\documentclass[showpacs,twocolumn,showkeys,amsmath,amssymb,pra,nofootinbib,subscriptaddress]{revtex4-1}

\usepackage{amsmath,amssymb,amsfonts}
\usepackage{graphicx}
\usepackage{txfonts}
\usepackage{epstopdf}
\usepackage[T1]{fontenc}

\usepackage[unicode]{hyperref}
\hypersetup{
   unicode=true,          
   a4paper=true,
   plainpages=false,
   pdftitle={Title of PDF},    
   pdfauthor={Author of PDF},     
   pdfsubject={Subject of PDF},   
   colorlinks=true,       
   citecolor=blue,        
}
\begin{document}

\title{From short-time diffusive to long-time ballistic dynamics: the unusual center-of-mass motion of quantum bright solitons
}

\author{Christoph Weiss}
\email{christoph.weiss@durham.ac.uk}
\affiliation{Joint Quantum Centre (JQC) Durham--Newcastle, Department of Physics, Durham University, Durham DH1 3LE, United Kingdom}

 \author{Simon A. Gardiner }
\affiliation{Joint Quantum Centre (JQC) Durham--Newcastle, Department of Physics, Durham University, Durham DH1 3LE, United Kingdom}

\author{Heinz-Peter Breuer}

\affiliation{Physikalisches Institut, Universit\"at Freiburg, Hermann-Herder-Stra{\ss}e 3, D-79104 Freiburg, Germany}

\date{\today}

 \begin{abstract}
Brownian motion is ballistic on short time scales and diffusive on long time scales. Our theoretical investigations indicate that one can observe the exact opposite --- an ``anomalous diffusion process'' where \textit{initially}\/  diffusive motion becomes ballistic on  \textit{longer}\/ time scales  ---  in an ultracold atom system with a size comparable to macromolecules. This system is the center-of-mass motion of a quantum matter-wave bright soliton for which the dominant  source of decoherence is \textit{three-particle losses}. Our simulations show that such unusual center-of-mass dynamics should be observable on experimentally accessible time scales.
\end{abstract}


\maketitle

\section{Introduction}

Bright solitons --- waves that do not change their shape --- were
discovered in the 19th century in a water
canal~\cite{Russell1845}. Such solitons are good examples of ballistic
motion (the distance from the initial position grows linearly with
time), as the velocity remains constant. Bright solitons can be
experimentally generated from attractively interacting ultracold
atomic
gases~\cite{KhaykovichEtAl2002,StreckerEtAl2002,CornishEtAl2006,MarchantEtAl2013,MedleyEtAl2014,McDonaldEtAl2014,NguyenEtAl2014};
on the mean-field level, via the Gross-Pitaevskii equation (GPE),
these matter-wave bright solitons are non-spreading solutions of a
non-linear
equation~\cite{PethickSmith2008,BaizakovEtAl2002,AlKhawajaEtAl2002,HaiEtAl2004,MartinRuostekoski2012,HelmEtAl2012,CuevasEtAl2013,PoloAhufinger2013}. For
$N$ ultracold attractively interacting atoms in a (quasi-)
one-dimensional wave guide,  \textit{quantum}\/ matter-wave
solitons~\cite{LaiHaus1989,DrummondEtAl1993,CarrBrand2004,StreltsovEtAl2011,FogartyEtAl2013,DelandeEtAl2013,GertjerenkenEtAl2013,BarbieroSalasnich2014}
can be described as a many-particle bound state. This is the ground state~\cite{McGuire1964,CalogeroDegasperis1975,CastinHerzog2001} of an exactly solvable many-particle quantum system, the Lieb-Liniger model~\cite{LiebLiniger1963} with attractive interactions~\cite{McGuire1964}. Already for particle numbers as low as three, these many-particle bound states share many similarities with mean-field matter-wave bright solitons~\cite{MazetsKurizki2006}.

Diffusive motion  [for which the root-mean-square (rms) fluctuations
of the position grow with the square root of time] of both
macromolecules and small classical particles often occurs through
interactions with the environment: Free Brownian
motion~\cite{UhlenbeckOrnstein1930,GrabertEtAl1988,JungHanggi1991,WangEtAl2002,LukicEtAl2005,KoepplEtAl2006,AndoSkolnick2010},
for example, exhibits the generic short-time-scale ballistic and long-time-scale diffusive behavior.  While there are models that,
depending on the choice of parameters, behave either diffusively
or ballistically~\cite{SteinigewegEtAl2007},  in this
 paper we show the surprising result that the dynamics of the rms fluctuations of the center of mass position of quantum bright
solitons, under the influence of decoherence via three-particle losses, behave diffusively on short
time scales and ballistically on long time scales.

Deviations from normal diffusion are an ongoing topic of current research. Anomalous diffusion~\cite{Metzler2000} has been observed experimentally in colloidal systems~\cite{SiemsEtAl2012,TurivEtAl2013}; research interest also includes superdiffusive motion~\cite{MetzlerKlafter2004}, which covers a regime in between diffusion and ballistic transport.  Diffusive and ballistic transport, and a surprising transition between the two are the focus of the current paper.

Diffusive behavior in
Bose-Einstein condensates has been observed in the experiment of Dries
et al.~\cite{DriesEtAl2010}, and for matter-wave bright solitons diffusive motion has been predicted in Ref.~\cite{SinhaEtAl2006}.
In this context it is important to note that, even for a perfect vacuum and when shielded from all external influence, decoherence via three-particle losses will always be present in an atomic Bose-Einstein condensate.  The only way to significantly decrease this source of decoherence would be to go to lower densities than is typical  for bright solitons as realized
experimentally, e.g., in
Refs.~\cite{KhaykovichEtAl2002,StreckerEtAl2002}. Thus, we focus on three-particle losses, which is for many parameter-regimes the dominant decoherence mechanism (cf.~\cite{WeissCastin2009}). For
 matter-wave bright solitons made of absolute ground-state atoms such as
\textsuperscript{7}Li~\cite{KhaykovichEtAl2002}, there are no two-particle
losses~\cite{GrimmEtAl2000}; single-particle losses can also be discounted if the
vacuum is made to be particularly good (cf.~\cite{AndersonEtAl1995}). It therefore is justified to focus on decoherence via three-particle losses.

The paper is organized as follows: We first introduce the physics involved in opening an initially weak trapping potential in which a bright soliton made from an attractive Bose-Einstein condensate has been prepared (Sec.~\ref{sec:open}). We then introduce the decoherence mechanism which will always be present in such a case --- atom losses via three-body recombination (Sec.~\ref{sec:three}), which is modeled via a stochastic approach using piecewise deterministic processes~\cite{Davis1993} in Sec.~\ref{sec:stochastic}. Section~\ref{sec:results} presents the results of our Monte Carlo simulation with the surprising transition from short-time diffusive to long-time ballistic behavior, and the paper ends with a conclusion and outlook (Sec.~\ref{sec:concl}).

\section{\label{sec:open}Opening a weak harmonic trap into a quasi-one dimensional wave guide}

\subsection{Mean field description: Stationary density profile}

When attractively interacting Bose-Einstein condensates are used experimentally to generate
bright solitons, the bright soliton is in a (quasi-)one
dimensional wave guide, that is, tight radial confinement and weak axial confinement~\cite{KhaykovichEtAl2002,StreckerEtAl2002,CornishEtAl2006,MarchantEtAl2013,MedleyEtAl2014,McDonaldEtAl2014}.
Important aspects of such bright solitons can be understood by the one-dimensional GPE~\cite{PethickSmith2008}
\begin{align}
\label{eq:GPE}
i\hbar \frac{\partial}{\partial t}\varphi = -\frac{\hbar^2}{2m}\frac{\partial^2}{\partial x^2}
\varphi +
 \frac {m\omega^2x^2}2\varphi
+g_{1 \rm D}(N-1)|\varphi|^2 \varphi,
\end{align}
where $m$ is the mass of the particles and $\omega$ the angular
frequency of the harmonic trap; the interaction $g_{\rm 1D}=2\hbar
\omega_{\perp}a$ is set by the
\textit{s}-wave scattering
length $a$ and the perpendicular angular trapping-frequency,
$\omega_{\perp}$~\cite{Olshanii1998}. 
For attractive interactions ($g_{\rm 1D}<0$) and weak harmonic trapping, Eq.~(\ref{eq:GPE}) has bright-soliton
solutions with single-particle densities $n\equiv |\varphi|^2$~\cite{PethickSmith2008}:
\begin{equation}
\label{eq:singlesoliton}
n(x)= \frac{1}{4\xi_N\left\{\cosh[x/(2\xi_N)]\right\}^2},
\end{equation}
where the soliton length is given by
\begin{equation}
\label{eq:solitonlength}
 \xi_N \equiv \frac{\hbar^2}{m\left|g_{\rm 1D}\right|(N-1)}.
\end{equation}
If the sufficiently weak (the soliton length~$\xi_N$ should be small compared to
the axial harmonic oscillator length~$\sqrt{\hbar/(m\omega)}$~\cite{Castin2009}) harmonic trap is then switched off, hardly any atoms are excited~\cite{Castin2009}. Thus, for bright solitons described on the mean-field (GPE) level, there will be no dynamics observable after opening the trap, whereas we will see that the same is not true for quantum bright solitons.

\subsection{ Quantum many-body description: Expansion of the
center-of-mass wave function}
In the absence of a trapping potential in the $x$-direction, the direction of the wave guide, all physically realistic $N$-particle models have to be translationally invariant in the $x$-direction [using the convention introduced in Eq.~(\ref{eq:GPE}) as the direction of the wave guide; $y$- and $z$-directions are harmonically trapped]. Thus, the center-of-mass eigenfunctions in the direction of the wave guide are plane waves and the center-of-mass dynamics resembles that of a heavy single particle, with the center-of-mass dynamics described by the Hamiltonian
\begin{equation}
\label{CoMHam}
 \hat{H} = -\frac{\hbar^2}{2Nm}\frac{\partial^2}{\partial X^2}
\end{equation}
and the center-of-mass coordinate given by the average of the positions of all $N$ particles
\begin{equation}
\label{eq:CoMCoord}
 X=\frac 1N\sum_{j=1}^Nx_j.
\end{equation}
The dynamics of the center of mass of an interacting gas in a
harmonic potential are independent of the interactions, giving
rise to the so-called ``Kohn mode'' \cite{BonitzEtAl2007}. Therefore, the initial center-of-mass wave function is independent of both the interactions and the approximate modeling of these interactions.

 Thus, the dynamics of the quantum bright soliton in the absence of potentials is due to the center-of-mass wave function of a particle of mass $M=Nm$~\cite{WeissCastin2009,Gertjerenken2013}. As the initial center-of-mass wave function is Gaussian, its time-dependence is~\cite{Fluegge1990}
\begin{align}
\label{eq:wavefunction}
\Psi(X,t) \propto &\left(1+i\frac{\hbar
    t}{2M\Delta X_0^2}\right)^{-1/2}\\
\nonumber &
\times \exp\left(-\frac{X^2-{i2\Delta X_0^2MV_0[X-V_0t]/\hbar}}{{4\Delta X_0^2}\left[1+i\hbar
      t/({2M\Delta X_0^2})\right]}\right),
\end{align}
where $X$ is the  center-of-mass coordinate~(\ref{eq:CoMCoord}) and $V_0$ the initial velocity. This implies an
rms width of~\cite{Fluegge1990}
\begin{equation}
\Delta X = \Delta X_0\sqrt{1+\left(\frac{\hbar t}{2M\Delta X_0^2}\right)^2}.
\label{eq:rms}
\end{equation}

For attractively interacting atoms ($g_{\rm 1D}<0$), the Lieb-Liniger-(McGuire) Hamiltonian~\cite{LiebLiniger1963,McGuire1964} is a very useful model
\begin{equation}
\hat{H} = -\sum_{j=1}^N\frac{\hbar^2}{2m}\frac{\partial^2}{\partial x_j^2}+\sum_{j=1}^{N-1}\sum_{n=j+1}^{N}g_{\rm 1D}\delta(x_j-x_n),
\label{eq:LL}
\end{equation}
where $x_j$ denotes the position of particle $j$. For this model, even the (internal) ground state wave function is known analytically. Including the center-of-mass momentum $K$, the corresponding eigenfunctions relevant for our dynamics read (cf.~\cite{CastinHerzog2001})
\begin{equation}
\Psi(x_1,x_2,\ldots,x_N) \propto e^{iKX} \exp\left(-\frac{m |g_{\rm 1D}|}{2\hbar^2}\sum_{j<\nu}|x_j-x_{\nu}|\right);
\end{equation}
the center-of-mass coordinate is given by Eq.~(\ref{eq:CoMCoord}). If the center-of-mass wave function is a delta function and the particle number is $N\gg 1$, then the single-particle density can be shown~\cite{CalogeroDegasperis1975,CastinHerzog2001} to be equivalent to the mean-field result~(\ref{eq:singlesoliton}). Thus, the Lieb-Liniger model is a one-dimensional many-particle quantum model that can be used to justify the approach to treat a quantum bright soliton like a mean-field soliton with additional center-of-mass motion after opening a weak initial trap. In the limit $N\to\infty$, $g_{\rm 1D}\to 0$ such that $Ng_{\rm 1D}= \rm const.$, the initial width of the center-of-mass wave function goes to zero, $\Delta X_0\propto 1/\sqrt{N}$.\footnote{Only for time scales $\propto \sqrt{N}$ does the width~(\ref{eq:rms}) of the center-of-mass wave function become visible when approaching the limit $N\to\infty$, $g_{\rm 1D}\to 0$ such that $Ng_{\rm 1D}= \rm const$. While such an agreement between GPE and $N$-particle quantum physics can be expected for some ground states~\cite{Lieb2002}, this is not necessarily true for many-particle dynamics~\cite{GertjerenkenWeiss2013}.}

\subsection{Single-particle density in the absence of decoherence}

Although the center-of-mass wave function~(\ref{eq:wavefunction}) spreads according to Eq.~(\ref{eq:rms}), 
 a single
measurement of the atomic density via scattering light off the soliton (cf.~\cite{KhaykovichEtAl2002}) will still yield the density profile of the soliton~(\ref{eq:singlesoliton}), expected both on the mean-field
(GPE) level and on the $N$-particle quantum level for
vanishing width of the center-of-mass wave
function~\cite{CalogeroDegasperis1975,CastinHerzog2001}. 
Taking into account
harmonic trapping perpendicular to the $x$-axis, one obtains the density~\cite{KhaykovichEtAl2002}
\begin{equation}
\label{eq:density}
n(x,y,z) = \frac N{4\xi_N\left\{\cosh\left[{x}/({2\xi_N})\right]\right\}^2}\frac1{\lambda_{\perp}^2{\pi}}\exp\left(-\frac{y^2+z^2}{\lambda_{\perp}^2}\right),
\end{equation}
where $\lambda_{\perp}\equiv\sqrt{\hbar/(m\omega_{\perp})}$ is the
perpendicular harmonic oscillator length.
In order to experimentally measure the spreading of the {center-of-mass
density} directly, each measurement of the soliton should only
record its center-of-mass position when calculating the density from the experimental data. Recording the entire density profile
in each measurement yields the single-particle density, which can also be obtained on a more formal level as a sum over the positions $\vec{x}_j$ of all particles
$
n(\vec{x})=\sum_{j=1}^N\langle\delta(\vec{x}-\vec{x}_j)\rangle/N
$.

\begin{figure}
\includegraphics[width=\linewidth]{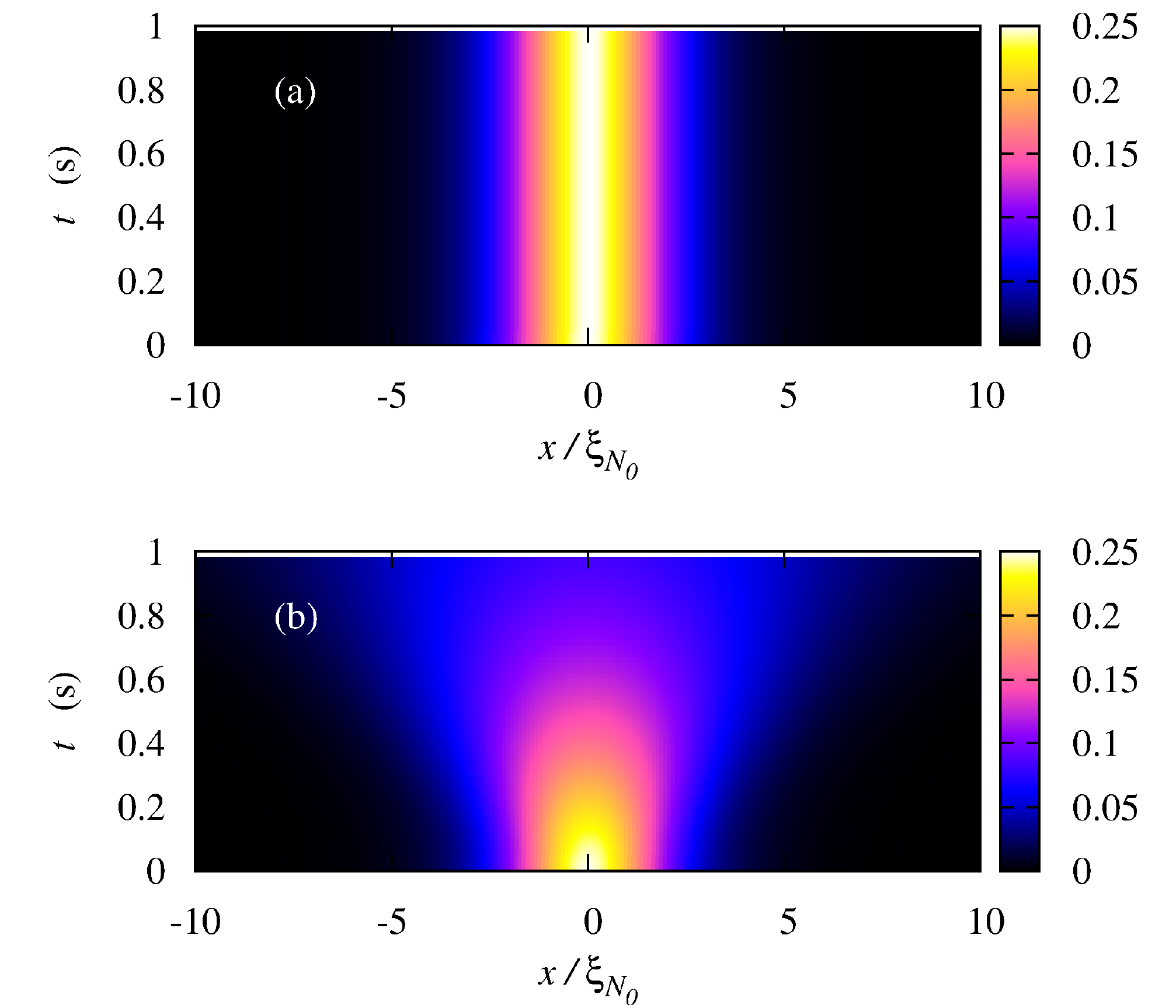}
\caption{\label{fig:transition}(Color online) 
Schematic display of the center-of-mass motion of a bright soliton without decoherence. 
Shown is the two-dimensional projection of the single-particle density
as a function of both time (measured in milliseconds) and distance from the
origin (measured in units of the initial soliton
width~$\xi_{N_0}$) for $N=3000$ Li atoms (using the parameters
of~\cite{KhaykovichEtAl2002}) after opening a weak initial harmonic
trap ($\lambda_{\rm HO}=10\xi_{N_0}$) at $t=0$. (a) On the
GPE level, the soliton remains stationary; the single particle density is given by Eq.~(\ref{eq:singlesoliton}). (b) On the many-particle quantum level, the ballistically expanding center-of-mass wave function smears out the single particle density. 
}
\end{figure}

Figure~\ref{fig:transition} shows the influence of the center-of-mass
position on the single particle density of a quantum bright soliton of
$3000$ Li atoms (as experimentally investigated at high velocities in~\cite{KhaykovichEtAl2002}). The GPE soliton remains stationary
[Fig.~\ref{fig:transition}~(a)]; that is, the single particle density is given by Eq.~(\ref{eq:singlesoliton}) for all times. Figure~\ref{fig:transition}~(b) displays the same situation as panel~(a) but for a quantum bright soliton for which the center-of-mass wave function spreads according to Eq.~(\ref{eq:rms}). Thus, for a quantum bright soliton we have a  spreading single particle density -- although each
single measurement yields the mean-field soliton density [Fig.~\ref{fig:transition}~(a)], shifted from
the initial position by some distance.

 For each single experiment, measuring the center-of-mass density of the many-particle configuration can be done with greater accuracy than the width of the cloud~(cf.~Ref.~\cite{GertjerenkenWeiss2012}). The expansion of the center-of-mass wave function leads to the spreading of the single particle density, in this paper we consider this spreading in the absence of harmonic trapping potentials. Recent experiments for homogeneous Bose gases can be found in Refs.~\cite{GauntEtAl2013,SchmidutzEtAl2014}.

\section{Decoherence via three-particle losses}

\subsection{Three particle-losses\label{sec:three}}

Three-particle losses can be described by a density-dependent rate equation \cite{GrimmEtAl2000}:
\begin{equation}
\frac{dN}{dt} = -K_3 \int d^3 r\, n^3(x,y,z),
\end{equation}
where $K_3$ is determined empirically.
Combined with Eq.~(\ref{eq:density}) and using the soliton length~(\ref{eq:solitonlength}) this yields~\cite{maple}
\begin{align}
\label{eq:dNdtmiddleright}
\frac{dN}{dt} &= -\frac{1}{90\pi^2}K_3\frac{1}{\xi_N^2\lambda_{\perp}^4}N^3 
= -\frac1{\tau_3} (N-1)^2N^3 ,
\end{align}
with the $N$-independent time scale:
\begin{equation}
{\tau_3}\equiv\frac {90\pi^2}{K_3 }\frac {\hbar^4\lambda_{\perp}^4}{m^2g_{\rm 1D}^2}.
\end{equation}
For this equation to be valid at longer time-scales (and not just
initially), the single particle density must remain of the
form~(\ref{eq:density}) as the width of the wave function in the $x$-direction increases with decreasing particle number. We will show that this assumption is self-consistent, thus allowing us to treat atom losses as point processes (referring to points in time) within our stochastic approach.

For large $N$, one may approximate Eq.~(\ref{eq:dNdtmiddleright}) by
$
{dN}/{dt} \simeq - N^5/\tau_3,
$
 which can be solved to give
\begin{align}
\label{eq:Ntleftright}
N(t) 
\simeq N_0\left(1 + \frac {t }{\tau_{\rm loss}}\right)^{-1/4},
\end{align}
where
\begin{equation}
\label{eq:Ntleftrightloss}
\tau_{\rm loss}
\equiv \frac{\tau_3}{4N_0^4}.
\end{equation}
For large initial particle numbers $N_0$ and experimentally relevant time scales
Eq.~(\ref{eq:Ntleftrightloss}) is a good approximation to the full time-dependence
[which will be shown in Fig.~\ref{fig:results}~(a)].

\subsection{\label{sec:stochastic}Stochastic modeling of decoherence via three-particle losses}

If changes to the number of particles in a soliton happen on slow enough time scales, these changes can be modeled as being adiabatic. The shape of the soliton is
protected~\cite{WeissCastin2009} (cf.~\cite{HelmEtAl2014}) by an energy gap
\begin{equation}
\label{eq:gap}
E_{\rm gap}(N) \equiv E_0(N-1)-E_0(N) =\frac{mg_{\rm 1D}^2N(N-1)}{8 \hbar^2},
\end{equation}
where $E_0(N) = -{mg_{1\rm D}^2}N(N^2-1)/(24{\hbar^2})$ is the ground state energy~\cite{McGuire1964} of a system of $N$ 1D point bosons of mass $m$ interacting via attractive delta interactions, described by the Lieb-Liniger Hamiltonian~(\ref{eq:LL}).

 The energy-time uncertainty yields a characteristic time scale (cf.~\cite{HoldawayEtAl2012}) via $E_{\rm gap}(N)\tau_{\rm soliton}(N)\propto \hbar$, where
\begin{equation}
\label{eq:tsoliton}
\tau_{\rm soliton}(N) = \frac{\hbar^3}{mg_{\rm 1D}^2(N-1)^2}.
\end{equation}
Changes in particle numbers should happen on time scales longer than this time for the process to be adiabatic, and for our approach of treating particle losses as an adiabatic process to be valid. 
So far, three-particle losses in experiments have not been
observed to destroy solitons on short time scales~\cite{KhaykovichEtAl2002}. We can thus model the
particle losses as taking place on time scales longer than the soliton
time if the soliton time is smaller than the time scale
$t_1 \simeq {\tau_3}/{N^5}$ on which a single particle is lost, that is,
\begin{equation}
\label{eq:ineq}
\frac{\tau_3}{N^5\tau_{\rm soliton}(N)} > 1.
\end{equation}
With $N_0=6000$ and the experimental parameters of
\cite{KhaykovichEtAl2002}\footnote{\label{footnote:parameters}The set of
  parameters used as an example to show that experimentally realistic
  time scales uses the values
  given in Ref.~\cite{KhaykovichEtAl2002} for the \textit{s}-wave scattering
  length~$a= -0.21 \times 10^{-9}\,\rm m$, $f_{\perp}=710\,\rm Hz$ where $\omega_{\perp} = 2\pi f_{\perp}$. 
For this \textit{s}-wave scattering length we furthermore divide the calculated
  value~\cite{ShotanEtAl2014} for the thermal $K_3$ of $3.6 \times
  10^{-41}\rm m^6/s$ by the factor $3! = 6$ for Bose-Einstein condensates
  and (thus also bright solitons).}
\begin{equation}
\label{eq:ratio}
\frac{\tau_3}{\tau_{\rm soliton}(N_0)}\approx2 \times 10^{20}.
\end{equation} 
The inequality~(\ref{eq:ineq}) is fulfilled for the parameters of~\cite{KhaykovichEtAl2002} if
$N  \lessapprox 6000$. We can furthermore model the three-particle losses as taking place instantaneously
for our stochastic implementation~\cite{DalibardEtAl1992,DumEtAl1992,Breuer2006} of
particle losses.

For a Schr\"odinger cat state~\cite{HarocheRaimond2006}, a quantum
superposition of two ``macroscopically'' occupied single particle
modes, $|\psi_{\rm NOON}\rangle \propto |1\rangle^{\otimes N} + 
|2\rangle^{\otimes N}$~\footnote{{The tensor product power notation $|1\rangle^{\otimes N}$ describes $N$ particles occupying the same single-particle mode $|1\rangle$.}}, losing three particles leads to a localization in one of
the two modes, $|1\rangle^{\otimes (N-3)}$ or $|2\rangle^{\otimes
  (N-3)}$. Quantum bright solitons are in a spatial quantum 
superposition given by their center-of-mass wave function; if the center-of-mass wave function is a
delta function, the wave function can be approximated by a
Hartree-product state consisting of occupying the mean-field
(GPE) wave function $N$ times~\cite{CastinHerzog2001}. We
thus will model the collapse {of the wave function} into one of these modes as a starting point
to describe the influence of decoherence via three-particle losses on the center-of-mass motion
of quantum bright solitons.

We can use the Schr\"odinger equation for a single particle of mass
$Nm$ with Hamiltonian
$ 
 \hat{H} = -[{\hbar^2}/({2Nm})]{\partial^2/\partial X^2}
$ 
 to describe the quantum mechanical motion of the center of mass~$X$  of a quantum bright soliton
in the absence of decoherence events~\cite{WeissCastin2009}.
Note that the particle number $N$
remains constant between loss events.

For the internal degrees of freedom, we can use a
Hartree-product-state~\cite{PethickSmith2008} of bright-soliton
solutions of the GPE~(\ref{eq:GPE}),
\begin{equation}
\label{eq:internal}
\psi_{V_0,N}(\underline{x}_N) =  \left\{\frac{e^{iNmV_0x/\hbar-i(\mu-Nmv^2/2) t/\hbar}}{2\sqrt{\xi_{N}}\cosh[(x-X_0+V_0t)/(2\xi_{N})]}\right\}^{\otimes N}
\end{equation}
with $\mu=g_{\rm 1D}(N-1)/(8\xi_N)$ and $\underline{x}_N=\{x_1,x_2.\ldots,x_N\}$.
After the three particle loss, the internal degrees of freedom are described by the wave function given by Eq.~(\ref{eq:internal}) with $N$ replaced by $N-3$.
As we will describe below,
both the position and the velocity (via the center-of-mass density) as
well as the point of time for
this decoherence [via Eq.~(\ref{eq:dNdtmiddleright})] are determined via random numbers in a
Monte Carlo simulation.
A characteristic size for the new center-of-mass wave function is the
root-mean-square width
of the soliton (cf.~Appendix~\ref{sec:appendix})
\begin{equation}
\label{eq:rmssoliton}
\Delta x_{\rm soliton} = \frac{\pi \xi_{N-3}}{\sqrt{3}}.
\end{equation}

In order to describe the stochastic process, we introduce an approach
via a classical master equation. While at first glance this approach
may seem to be impossible, as between loss events we have a purely quantum mechanical expansion of the center-of-mass wave function,  the fact that our system can indeed be described by a classical model is justified {below}.
Within our model the stochastic variables are given by the center of mass
coordinate $X$, the corresponding velocity $V$ and the particle number $N$.
Introducing the time-dependent probability distribution $P(X,V,N,t)$ the stochastic
process is defined by the master equation
\begin{align}
  \frac{\partial}{\partial t} &P(X,V,N,t) = -V\frac{\partial}{\partial X} P(X,V,N,t) \nonumber \\
 + &  \int dX' \int dV' \left[ W_{N+3}(X,V|X',V')P(X',V',N+3,t)\right.  \nonumber \\ 
 &- \left. W_{N}(X',V'|X,V)P(X,V,N,t) \right]. 
\end{align}
The first term on the right-hand side describes the constant drift of $X$ with
velocity $V$ while the second term represents the instantaneous random jumps
induced by three-particle losses. The process $(X,V,N)$ is thus a
piecewise deterministic process~\cite{Davis1993} with transition rates 
\begin{align}
 W_{N}(X',V'|X,V) = \Gamma(N) &\sqrt{\frac{1}{2\pi\sigma^2_X(N)}} 
\exp\left({-\frac{(X-X')^2}{2\sigma^2_X(N)}}\right)\nonumber\\ \times& \sqrt{\frac{1}{2\pi\sigma^2_V(N)}} 
\exp\left({-\frac{(V-V')^2}{2\sigma^2_V(N)}}\right),
\end{align}
where $\sigma_X(N)$ is given by Eq.~(\ref{eq:rmssoliton}), and 
\[\sigma_V(N) = \hbar/[2m(N-3)\sigma_X(N)].\] 
The total transition rate takes the form
\begin{equation}
\label{eq:Gamma}
 \Gamma(N) \equiv \int dX' \int dV' W_{N}(X',V'|X,V) = \frac{(N-1)^2N^3}{3\tau_3},
\end{equation}
where we have added a factor of $1/3$ as three particles are lost each time. We thus have an exponential waiting time distribution 
\begin{equation}
\label{eq:losstimes}
 F(N,t) = 1 - \exp\left[-{\Gamma(N)} t\right] .
\end{equation}


To summarize, for the stochastic simulation of decoherence
via three-particle losses~\cite{Breuer2006}, the ingredients are:

\begin{enumerate}
\item
The random variables:
\begin{equation}
N, X_0, V_0
\end{equation}
\item Random numbers for the Monte-Carlo process determine:
\begin{enumerate}
\item The time of the next decoherence event via Eq.~(\ref{eq:dNdtmiddleright})
  by choosing an exponential distribution of loss times~(\ref{eq:losstimes}),
  where the factor $1/3$ introduced in Eq.~(\ref{eq:Gamma}) is necessary because \textit{three}\/ particles are lost in
  each step: $N\rightarrow N-3$.
\item The center-of-mass position $X_0$ of the new
  wave function via the center-of-mass density in
  real space.
\item The  center-of-mass velocity $V_0$ of the new
  wave function via the center-of-mass density in
  momentum space.
\end{enumerate}
\item The center-of-mass wave function corresponding to the product
  state~(\ref{eq:internal}) is chosen to be a Gaussian
\begin{equation}
\psi_{\rm CoM} = \exp\left[-\frac{(X-X_0)^2}{2b^2}+i\frac{(N-3)mV_0}{\hbar}X\right]
\end{equation}
with a root-mean-square width $\sigma_X(N)=b/\sqrt{2}$ given
\label{method2} by  Eq.~(\ref{eq:rmssoliton})
\begin{equation}
\label{eq:rms2}
\sigma_X(N)= \frac{\pi \xi_{N-3}}{\sqrt{3}}.
\end{equation}
\end{enumerate}


In between loss events, the quantum dynamics is known analytically
[Eq.~(\ref{eq:wavefunction})]; rather than solving the Schr\"odinger equation it is possible to
do this in a more classical approach: The truncated Wigner
approximation~\footnote{The truncated-Wigner approximation~\cite{SinatraEtAl2002} describes quantum systems by averaging over realizations of an
appropriate classical field equation (in this case, the GPE) with initial noise
appropriate to either finite~\cite{BieniasEtAl2011} or zero
temperatures~\cite{MartinRuostekoski2012}. } for the center of mass,
which has been used in Ref.~\cite{GertjerenkenEtAl2013} to qualitatively
mimic quantum behavior on the mean-field level by introducing
classical noise mimicking the quantum uncertainties in both position
and momentum, is particularly useful here: both the mean position and
the variance calculated via the Truncated Wigner Approximation for the center of mass are identical to the quantum mechanical result. In order to make both results identical, Gaussian noise has to be added independently to
both position $X_0\rightarrow X= X_0 + \delta X_0$ and velocity
$V_0\rightarrow V=V_0+ \delta V_0$ with $\langle \delta X_0\rangle =0$ and
$\langle \delta V_0\rangle =0$ and root-mean-square fluctuations
$\sigma_X(N)$ given
by  Eq.~(\ref{eq:rms2}) and by the
minimal uncertainty relation
\begin{equation}
\sigma_V(N)=\frac{\hbar}{2(N-3)m \sigma_X(N)}
\end{equation}
for the velocity.

 The mean position $\overline{x(t)}=\overline{X_0+V_0t}$ is thus
 identical to the quantum mechanical result; the root-mean-square
 fluctuations $\Delta x = \sqrt{(\Delta X_0)^2+(\Delta V_0)^2t^2}$
 coincide with the quantum mechanical equation~(\ref{eq:rms}). Thus,
 in the absence of both the trap in the axial direction and the
 scattering processes investigated in
 Ref.~\cite{GertjerenkenEtAl2013}, the TWA for the center of mass gives exact results for both the position of the center of mass and the root-mean-square fluctuations of the center of mass for a quantum bright soliton.

\section{\label{sec:results}Results}

\begin{figure}
\includegraphics[width=\linewidth]{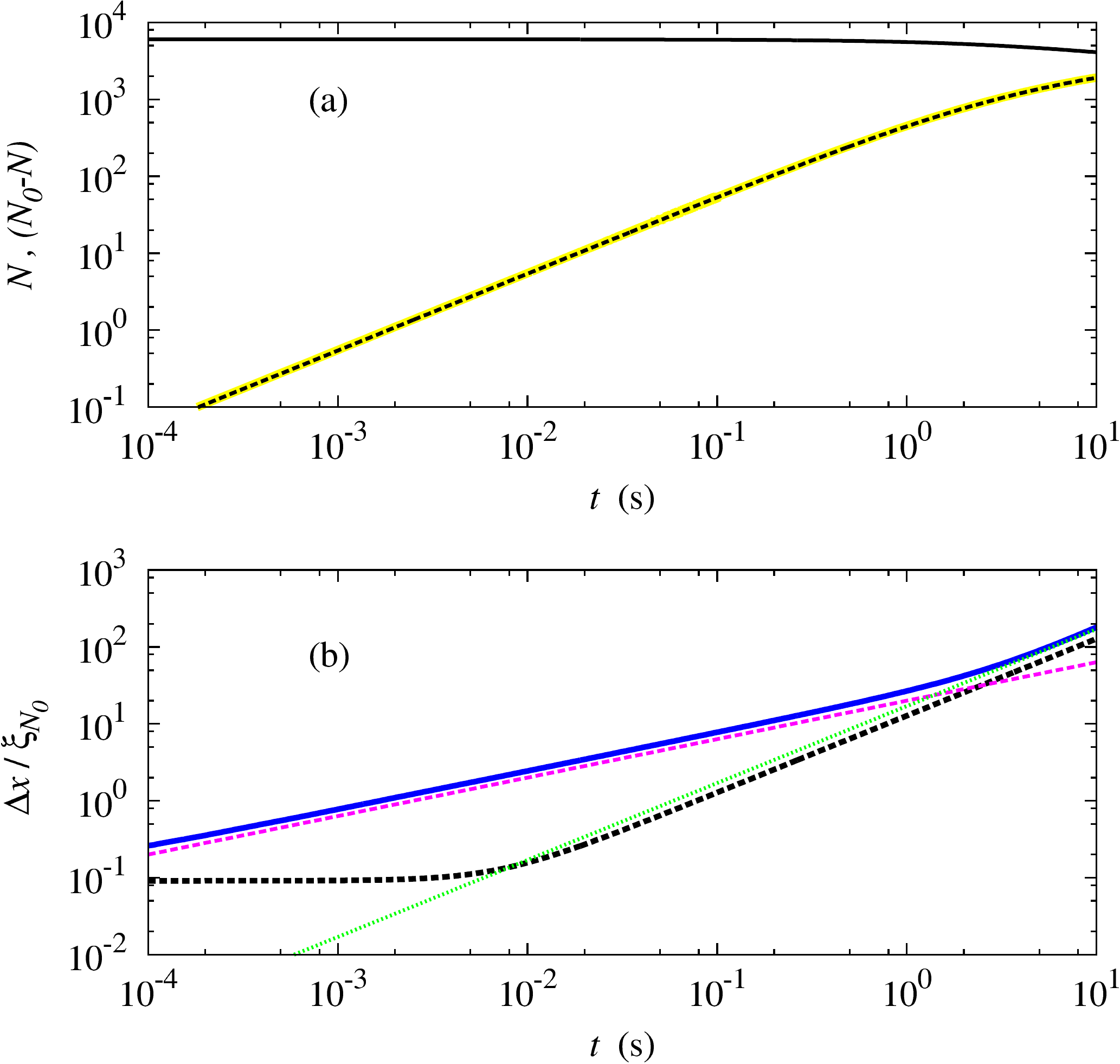}
\caption{\label{fig:results}(Color online)  Influence of decoherence via three-particle losses on particle number and rms width of the single-particle density using the parameters described in  footnote~\ref{footnote:parameters}.
 (a) Number of particles $N(t)$  obtained numerically as a function of time (solid black
line) with $N(0)=6000$.  For $6000-N(t)$ as a function of time, the numerical curve (black dashed
line) lies on top of the
analytic curve [yellow/light gray, Eq.~(\ref{eq:Ntleftright})]. (b) As soon as particle losses [modeled via the piecewise deterministic processes described in Sec.~\ref{sec:stochastic} using Eqs.~(\ref{eq:ratio}) and (\ref{eq:rms2})] become important, the
rms width of the single-particle density of a bright
soliton (blue/black solid line) grows like the
square-root of time (cf.~dashed magenta/dark gray line) before becoming
ballistic at larger times [$\propto t$, green/light gray line], approaching the width of the center of mass wave function without decoherence (black dashed line) for larger times.}
\end{figure}

Figure~\ref{fig:results} shows the influence of decoherence via three-particle
losses on the center-of-mass displacement of quantum bright solitons made out of Li atoms for parameters taken from the experiment~\cite{KhaykovichEtAl2002} (see footnote~\ref{footnote:parameters}). Three-particle losses, which could only be prevented by considerably reducing the density of a bright soliton to values much lower than used in experiments such as~\cite{KhaykovichEtAl2002}, and are thus a decoherence mechanism intrinsic to quantum bright solitons, lead to a transition from short-time diffusive to long-time ballistic behavior [Fig.~\ref{fig:results}~(b)]. The numerical simulations were done by using the piece-wise deterministic processes~\cite{Davis1993} described in Sec.~\ref{sec:stochastic}, a well-established tool to model decoherence~\cite{DalibardEtAl1992,DumEtAl1992,Breuer2006}.

\begin{figure}
\includegraphics[width=\linewidth]{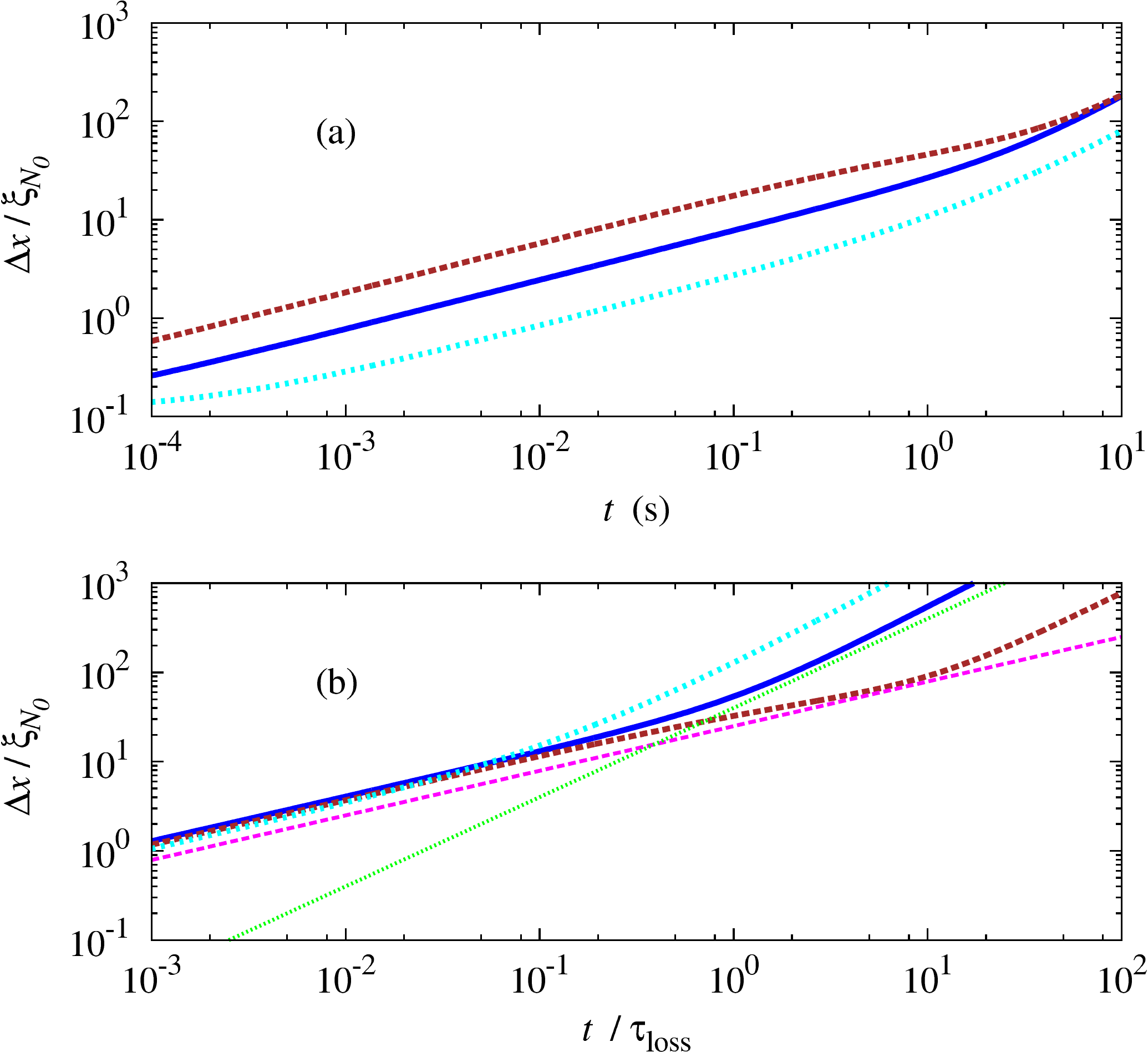}
\caption{\label{fig:scaling1} {(Color online) (a) Root-mean-square (rms) width of center-of-mass displacement as a function of time. (a) The data shown is for:
blue/black solid line: $N_0=6000$, $\tau_3/\tau_{\rm soliton}=2 \times
10^{20}$, light blue/light gray dashed line: $N_0=4000$,
$\tau_3/\tau_{\rm soliton}=1 \times 10^{20}$, brown/black dashed line:
$N_0=5000$, $\tau_3/\tau_{\rm soliton}=1 \times 10^{19}$. (b) Same
data sets as in the previous panel but with a time-axis rescaled with
the characteristic time from Eq.~(\ref{eq:Ntleftrightloss}). Both the
magenta/dark gray straight line: $\propto \sqrt{t}$, and the
green/light gray straight line $\propto {t}$ where added as guides to
the eye.}}
\end{figure}
Figure~\ref{fig:scaling1} shows that the transition from short-time  diffusive to
long-time ballistic behavior is not dependent on a particular choice of parameters. While details of the curves can look different for different parameters, the transition from short-term diffusive to long-time ballistic behavior is visible in particular after rescaling the time-axis with the characteristic time-scale given by the atom losses [Eq.~(\ref{eq:Ntleftright})]; thus
using the scaling for which all curves $N(t)/N_0$ would lie on top of
each other. Within this scaling, all curves follow the same $\sqrt{t/\tau_{\rm loss}}$ scaling for \emph{early} scaled times,  and they each start deviating off at \emph{different} times.

This scaling leads to an intuitive explanation of the transition from short-time diffusive to long-time ballistic motion. While the atom losses continue in the regime of ballistic motion [as can be seen by comparing panels (a) and (b) of Fig.~\ref{fig:results}], $\tau_{\rm loss}$ is the time-scale on which $N(t)$ starts to forget its initial number of particles. In addition, the center-of-mass motion also picks up pace for longer time-scales.
 As the Gross-Pitaevskii equation (GPE) becomes valid in the limit 
$N\to\infty$, $g_{\rm 1D}\to 0$ such that the product $Ng_{\rm 1D}$ 
remains constant~\cite{LiebEtAl2000}, it cannot model an expanding center-of-mass wave function. Thus, the transition from short-time ballistic to long-time diffusive behavior cannot be modeled by simply using standard GPE-theory.

\section{\label{sec:concl}Conclusion and outlook}

To conclude,
we have introduced a physically motivated model for the motion of quantum bright solitons which displays  short-time diffusive and long-time ballistic behavior, contrary to the usual short-time ballistic and long-time diffusive behavior observed for example in Brownian motion~\cite{LukicEtAl2005}. Bright solitons are investigated experimentally in various groups world-wide. As the ballistic expansion for large times is $\propto t/(Nm)$ [Eq.~(\ref{eq:rms})] the solitons made of thousands of Li atoms~\cite{KhaykovichEtAl2002} are more suitable to observe this motion of the center-of-mass than solitons made of thousands of the more than ten times heavier Rb atoms~\cite{MarchantEtAl2013}.  For the ground-state atoms of Li used, for example,  in the ground-breaking experiments~\cite{KhaykovichEtAl2002,StreckerEtAl2002} there are no two-body losses~\cite{GrimmEtAl2000}; single-particle losses can also be discounted if the vacuum is made to be particularly good (cf.~\cite{AndersonEtAl1995}). Our approach to focus on decoherence via three-particle losses to model matter-wave bright solitons in attractive Li-Bose-Einstein condensates thus is justified.

The present idea to modify the quantum mechanical motion by
stochastic terms in order to describe instantaneous changes
of the wave function to smaller wave packets has formal
similarities with stochastic collapse
models~\cite{GhirardiEtAl1986}. However, within our model these random changes
describe the decoherence of the center-of-mass wave
function which is induced by three-particle losses;  a decoherence mechanism which cannot be avoided by, e.g., choosing a perfect vacuum: as long as the density is finite (which always is the case for bright solitons),  three-particle losses will occur as a dominant decoherence mechanism. It is
the decrease of the particle number that leads to fewer particle losses and, hence, to the
observed transition from diffusive to ballistic motion. 
This motion is an effect distinct from both classical~\cite{Metzler2000} and quantum walks cf.~\cite{DurEtAl2002,KarskiEtAl2009} as well as anomalous diffusion~\cite{Metzler2000,SiemsEtAl2012,TurivEtAl2013}. As for the classical random walk, our model localizes after each step, but between steps the motion is given by free expansion of the center-of-mass wave function which depends on the (decreasing) number of particles.

This unusual behavior of the center-of-mass motion can be observed for experimentally realistic parameters; both time scales and length scales are accessible experimentally.

\acknowledgments

We thank S.\ L.\ Cornish, S.\ A.\ Hopkins and L.\ Khaykovich for
discussions. C.W.\ thanks the Institute of Physics, University of Freiburg, for its hospitality.  We thank the UK Engineering and Physical Sciences Research Council (Grant No.\ EP/L010844/1, C.W.\ and
S.A.G.) for funding. 
The data presented in this paper
are available
 from~\url{http://dx.doi.org/10.15128/kk91fk954}.

\begin{appendix}

\section{\label{sec:appendix}Size of Center-of-Mass wave function after collapse}

\begin{figure}
\includegraphics[width=\linewidth]{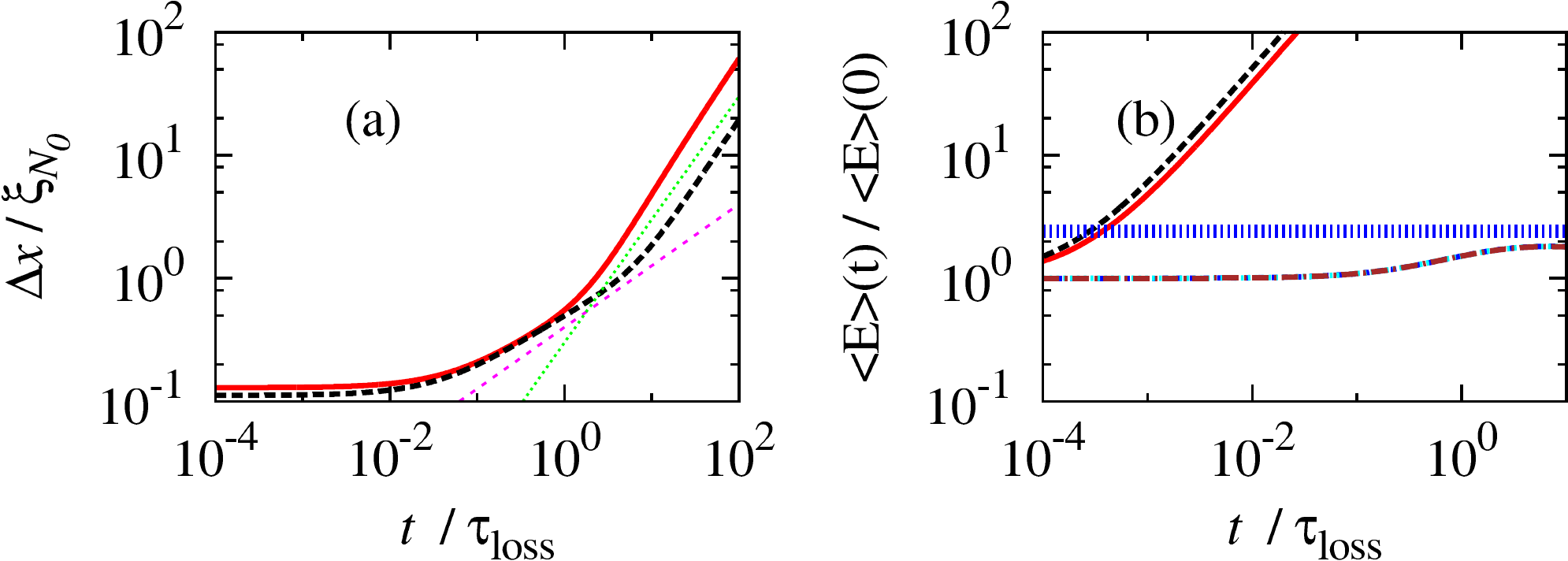}
\caption{\label{fig:appendix}{Root-mean-square (rms) width of center-of-mass displacement as a function of time for a model differing from the choice in the main part of the paper (Figs.~\ref{fig:results} and \ref{fig:scaling1}). Solid red/black curve: $N_0=3000$, $\tau_3/\tau_{\rm soliton}=2 \times 10^{15}$; black dashed curve:  $N_0=4000$, $\tau_3/\tau_{\rm soliton}=2 \times 10^{15}$. Dashed magenta/dark gray line $\propto \sqrt{t}$; green/light gray line $\propto t$.  (b)  The kinetic energy as a function of time grows considerably for the two curves of panel (a) whereas it stays below the line corresponding to $(\langle E\rangle(0) + \Delta E(0))/\langle E\rangle(0)$ (thick blue/black dashed horizontal line) for the curves of Fig.~\ref{fig:scaling1}.}}
\end{figure}

The focus of this paper was to present a physically motivated model
which displays a transition from short-time diffusive to long-time
ballistic behavior. Time-scales can easily be changed by, e.g.,
choosing a trapping geometry different from the parameters used in
Ref.~\cite{KhaykovichEtAl2002}. The focus currently is on a
macroscopic theory; for future microscopic theories some details like
the center-of-mass wave function after a decoherence-event via the physically dominating decoherence mechanism, a three-particle loss-event, might differ from the value chosen here. In order to show that the transition from diffusive to ballistic behavior would still be observable for other choices of the width of the center-of-mass wave function, Fig.~\ref{fig:appendix} displays the behavior for 
\begin{equation}
\label{eq:appendix}
\Delta X_{\rm CoM} = \frac{\pi \xi_{N-3}}{\sqrt{3(N-3)}}.
\end{equation}
This corresponds to the idealized case that the wave function collapses to a single product state~(\ref{eq:internal}), the root-mean-square width of the new center-of-mass wave
  function of the soliton consisting of $N-3$ particles is given by
  the prediction of the central limit theorem (cf.~\cite{HoldawayEtAl2012})

Figure~\ref{fig:appendix} shows, that 
as for the choice in the main part of the paper, for Eq.~(\ref{eq:appendix}) the combined effect of the rate of particle losses decreasing and becoming more independent of $N(t=0)$ [cf.~Eq.~(\ref{eq:Ntleftright})] and the center-of-mass motion covering greater distances leads again to a transition from short-time diffusive to long-time ballistic behavior. However, contrary to the case discussed in the  main part of the paper, the kinetic energy is considerably increased during the motion. While the open system discussed in this paper could include such a mechanism, unless experimental results should oblige one to introduce such a mechanism, the model presented in the main part of the paper is the more physical choice.

\end{appendix}
%


\begin{thebibliography}{10}%
\makeatletter
\providecommand \@ifxundefined [1]{%
 \ifx #1\undefined \expandafter \@firstoftwo
 \else \expandafter \@secondoftwo
\fi
}%
\providecommand \@ifnum [1]{%
 \ifnum #1\expandafter \@firstoftwo
 \else \expandafter \@secondoftwo
\fi
}%
\providecommand \enquote [1]{``#1''}%
\providecommand \bibnamefont  [1]{#1}%
\providecommand \bibfnamefont [1]{#1}%
\providecommand \citenamefont [1]{#1}%
\providecommand\href[0]{\@sanitize\@href}%
\providecommand\@href[1]{\endgroup\@@startlink{#1}\endgroup\@@href}%
\providecommand\@@href[1]{#1\@@endlink}%
\providecommand \@sanitize [0]{\begingroup\catcode`\&12\catcode`\#12\relax}%
\@ifxundefined \pdfoutput {\@firstoftwo}{%
 \@ifnum{\z@=\pdfoutput}{\@firstoftwo}{\@secondoftwo}%
}{%
 \providecommand\@@startlink[1]{\leavevmode}%
 \providecommand\@@endlink[0]{}%
}{%
 \providecommand\@@startlink[1]{%
  \leavevmode
  \pdfstartlink
   attr{/Border[0 0 1 ]/H/I/C[0 1 1]}%
   user{/Subtype/Link/A<</Type/Action/S/URI/URI(#1)>>}%
  \relax
 }%
 \providecommand\@@endlink[0]{\pdfendlink}%
}%
\providecommand \url  [0]{\begingroup\@sanitize \@url }%
\providecommand \@url [1]{\endgroup\@href {#1}{\urlprefix}}%
\providecommand \urlprefix [0]{URL }%
\providecommand \Eprint[0]{\href }%
\@ifxundefined \urlstyle {%
  \providecommand \doi [1]{doi:\discretionary{}{}{}#1}%
}{%
  \providecommand \doi [0]{doi:\discretionary{}{}{}\begingroup
  \urlstyle{rm}\Url }%
}%
\providecommand \doibase [0]{http://dx.doi.org/}%
\providecommand \Doi[1]{\href{\doibase#1}}%
\providecommand \bibAnnote [3]{%
  \BibitemShut{#1}%
  \begin{quotation}\noindent
    \textsc{Key:}\ #2\\\textsc{Annotation:}\ #3%
  \end{quotation}%
}%
\providecommand \bibAnnoteFile [2]{%
  \IfFileExists{#2}{\bibAnnote {#1} {#2} {\input{#2}}}{}%
}%
\providecommand \typeout [0]{\immediate \write \m@ne }%
\providecommand \selectlanguage [0]{\@gobble}%
\providecommand \bibinfo [0]{\@secondoftwo}%
\providecommand \bibfield [0]{\@secondoftwo}%
\providecommand \translation [1]{[#1]}%
\providecommand \BibitemOpen[0]{}%
\providecommand \bibitemStop [0]{}%
\providecommand \bibitemNoStop [0]{.\EOS\space}%
\providecommand \EOS [0]{\spacefactor3000\relax}%
\providecommand \BibitemShut [1]{\csname bibitem#1\endcsname}%
\bibitem{Russell1845}%
  \BibitemOpen
  \bibfield{author}{%
  \bibinfo {author} {\bibfnamefont{J.~S.}\ \bibnamefont{Russell}},\ }%
  \bibfield{journal}{%
  \bibinfo {journal} {\textrm{In} Report of the Fourteenth Meeting of the
  British Association for the Advancement of Science 311--390 (John Murray)}}%
   (\bibinfo {year} {1845})%
  \bibAnnoteFile{NoStop}{Russell1845}%
\bibitem{KhaykovichEtAl2002}%
  \BibitemOpen
  \bibfield{author}{%
  \bibinfo {author} {\bibfnamefont{L.}~\bibnamefont{Khaykovich}}, \bibinfo
  {author} {\bibfnamefont{F.}~\bibnamefont{Schreck}}, \bibinfo {author}
  {\bibfnamefont{G.}~\bibnamefont{Ferrari}}, \bibinfo {author}
  {\bibfnamefont{T.}~\bibnamefont{Bourdel}}, \bibinfo {author}
  {\bibfnamefont{J.}~\bibnamefont{Cubizolles}}, \bibinfo {author}
  {\bibfnamefont{L.~D.}\ \bibnamefont{Carr}}, \bibinfo {author}
  {\bibfnamefont{Y.}~\bibnamefont{Castin}},\ and\ \bibinfo {author}
  {\bibfnamefont{C.}~\bibnamefont{Salomon}},\ }%
  \bibfield{journal}{%
  \Doi{10.1126/science.1071021}{\bibinfo {journal} {Science}}\ }%
  \textbf{\bibinfo {volume} {296}},\ \bibinfo {pages} {1290} (\bibinfo {year}
  {2002})%
  \bibAnnoteFile{NoStop}{KhaykovichEtAl2002}%
\bibitem{StreckerEtAl2002}%
  \BibitemOpen
  \bibfield{author}{%
  \bibinfo {author} {\bibfnamefont{K.~E.}\ \bibnamefont{Strecker}}, \bibinfo
  {author} {\bibfnamefont{G.~B.}\ \bibnamefont{Partridge}}, \bibinfo {author}
  {\bibfnamefont{A.~G.}\ \bibnamefont{Truscott}},\ and\ \bibinfo {author}
  {\bibfnamefont{R.~G.}\ \bibnamefont{Hulet}},\ }%
  \bibfield{journal}{%
  \Doi{10.1038/nature747}{\bibinfo {journal} {Nature (London)}}\ }%
  \textbf{\bibinfo {volume} {417}},\ \bibinfo {pages} {150} (\bibinfo {year}
  {2002})%
  \bibAnnoteFile{NoStop}{StreckerEtAl2002}%
\bibitem{CornishEtAl2006}%
  \BibitemOpen
  \bibfield{author}{%
  \bibinfo {author} {\bibfnamefont{S.~L.}\ \bibnamefont{Cornish}}, \bibinfo
  {author} {\bibfnamefont{S.~T.}\ \bibnamefont{Thompson}},\ and\ \bibinfo
  {author} {\bibfnamefont{C.~E.}\ \bibnamefont{Wieman}},\ }%
  \bibfield{journal}{%
  \Doi{10.1103/PhysRevLett.96.170401}{\bibinfo {journal} {Phys. Rev. Lett.}}\
  }%
  \textbf{\bibinfo {volume} {96}},\ \bibinfo {pages} {170401} (\bibinfo {year}
  {2006})%
  \bibAnnoteFile{NoStop}{CornishEtAl2006}%
\bibitem{MarchantEtAl2013}%
  \BibitemOpen
  \bibfield{author}{%
  \bibinfo {author} {\bibfnamefont{A.~L.}\ \bibnamefont{{Marchant}}}, \bibinfo
  {author} {\bibfnamefont{T.~P.}\ \bibnamefont{{Billam}}}, \bibinfo {author}
  {\bibfnamefont{T.~P.}\ \bibnamefont{{Wiles}}}, \bibinfo {author}
  {\bibfnamefont{M.~M.~H.}\ \bibnamefont{{Yu}}}, \bibinfo {author}
  {\bibfnamefont{S.~A.}\ \bibnamefont{{Gardiner}}},\ and\ \bibinfo {author}
  {\bibfnamefont{S.~L.}\ \bibnamefont{{Cornish}}},\ }%
  \bibfield{journal}{%
  \Doi{10.1038/ncomms2893}{\bibinfo {journal} {Nat. Commun.}}\ }%
  \textbf{\bibinfo {volume} {4}},\ \bibinfo {pages} {1865} (\bibinfo {year}
  {2013})%
  \bibAnnoteFile{NoStop}{MarchantEtAl2013}%
\bibitem{MedleyEtAl2014}%
  \BibitemOpen
  \bibfield{author}{%
  \bibinfo {author} {\bibfnamefont{P.}~\bibnamefont{Medley}}, \bibinfo {author}
  {\bibfnamefont{M.~A.}\ \bibnamefont{Minar}}, \bibinfo {author}
  {\bibfnamefont{N.~C.}\ \bibnamefont{Cizek}}, \bibinfo {author}
  {\bibfnamefont{D.}~\bibnamefont{Berryrieser}},\ and\ \bibinfo {author}
  {\bibfnamefont{M.~A.}\ \bibnamefont{Kasevich}},\ }%
  \bibfield{journal}{%
  \Doi{10.1103/PhysRevLett.112.060401}{\bibinfo {journal} {Phys. Rev. Lett.}}\
  }%
  \textbf{\bibinfo {volume} {112}},\ \bibinfo {pages} {060401} (\bibinfo {year}
  {2014})%
  \bibAnnoteFile{NoStop}{MedleyEtAl2014}%
\bibitem{McDonaldEtAl2014}%
  \BibitemOpen
  \bibfield{author}{%
  \bibinfo {author} {\bibfnamefont{G.~D.}\ \bibnamefont{McDonald}}, \bibinfo
  {author} {\bibfnamefont{C.~C.~N.}\ \bibnamefont{Kuhn}}, \bibinfo {author}
  {\bibfnamefont{K.~S.}\ \bibnamefont{Hardman}}, \bibinfo {author}
  {\bibfnamefont{S.}~\bibnamefont{Bennetts}}, \bibinfo {author}
  {\bibfnamefont{P.~J.}\ \bibnamefont{Everitt}}, \bibinfo {author}
  {\bibfnamefont{P.~A.}\ \bibnamefont{Altin}}, \bibinfo {author}
  {\bibfnamefont{J.~E.}\ \bibnamefont{Debs}}, \bibinfo {author}
  {\bibfnamefont{J.~D.}\ \bibnamefont{Close}},\ and\ \bibinfo {author}
  {\bibfnamefont{N.~P.}\ \bibnamefont{Robins}},\ }%
  \bibfield{journal}{%
  \Doi{10.1103/PhysRevLett.113.013002}{\bibinfo {journal} {Phys. Rev. Lett.}}\
  }%
  \textbf{\bibinfo {volume} {113}},\ \bibinfo {pages} {013002} (\bibinfo {year}
  {2014})%
  \bibAnnoteFile{NoStop}{McDonaldEtAl2014}%
\bibitem{NguyenEtAl2014}%
  \BibitemOpen
  \bibfield{author}{%
  \bibinfo {author} {\bibfnamefont{J.~H.~V.}\ \bibnamefont{{Nguyen}}}, \bibinfo
  {author} {\bibfnamefont{P.}~\bibnamefont{{Dyke}}}, \bibinfo {author}
  {\bibfnamefont{D.}~\bibnamefont{{Luo}}}, \bibinfo {author}
  {\bibfnamefont{B.~A.}\ \bibnamefont{{Malomed}}},\ and\ \bibinfo {author}
  {\bibfnamefont{R.~G.}\ \bibnamefont{{Hulet}}},\ }%
  \bibfield{journal}{%
  \Doi{10.1038/nphys3135}{\bibinfo {journal} {Nature Phys.}}\ }%
  \textbf{\bibinfo {volume} {10}},\ \bibinfo {pages} {918} (\bibinfo {year}
  {2014})%
  \bibAnnoteFile{NoStop}{NguyenEtAl2014}%
\bibitem{PethickSmith2008}%
  \BibitemOpen
  \bibfield{author}{%
  \bibinfo {author} {\bibfnamefont{C.~J.}\ \bibnamefont{Pethick}}\ and\
  \bibinfo {author} {\bibfnamefont{H.}~\bibnamefont{Smith}},\ }%
  \emph{\bibinfo {title} {Bose-Einstein Condensation in Dilute Gases}}\
  (\bibinfo {publisher} {Cambridge University Press},\ \bibinfo {address}
  {Cambridge},\ \bibinfo {year} {2008})%
  \bibAnnoteFile{NoStop}{PethickSmith2008}%
\bibitem{BaizakovEtAl2002}%
  \BibitemOpen
  \bibfield{author}{%
  \bibinfo {author} {\bibfnamefont{B.~B.}\ \bibnamefont{Baizakov}}, \bibinfo
  {author} {\bibfnamefont{V.~V.}\ \bibnamefont{Konotop}},\ and\ \bibinfo
  {author} {\bibfnamefont{M.}~\bibnamefont{Salerno}},\ }%
  \bibfield{journal}{%
  \Doi{10.1088/0953-4075/35/24/312}{\bibinfo {journal} {J. Phys. B}}\ }%
  \textbf{\bibinfo {volume} {35}},\ \bibinfo {pages} {5105} (\bibinfo {year}
  {2002})%
  \bibAnnoteFile{NoStop}{BaizakovEtAl2002}%
\bibitem{AlKhawajaEtAl2002}%
  \BibitemOpen
  \bibfield{author}{%
  \bibinfo {author} {\bibfnamefont{U.}~\bibnamefont{Al~Khawaja}}, \bibinfo
  {author} {\bibfnamefont{H.~T.~C.}\ \bibnamefont{Stoof}}, \bibinfo {author}
  {\bibfnamefont{R.~G.}\ \bibnamefont{Hulet}}, \bibinfo {author}
  {\bibfnamefont{K.~E.}\ \bibnamefont{Strecker}},\ and\ \bibinfo {author}
  {\bibfnamefont{G.~B.}\ \bibnamefont{Partridge}},\ }%
  \bibfield{journal}{%
  \Doi{10.1103/PhysRevLett.89.200404}{\bibinfo {journal} {Phys. Rev. Lett.}}\
  }%
  \textbf{\bibinfo {volume} {89}},\ \bibinfo {pages} {200404} (\bibinfo {year}
  {2002})%
  \bibAnnoteFile{NoStop}{AlKhawajaEtAl2002}%
\bibitem{HaiEtAl2004}%
  \BibitemOpen
  \bibfield{author}{%
  \bibinfo {author} {\bibfnamefont{W.}~\bibnamefont{Hai}}, \bibinfo {author}
  {\bibfnamefont{C.}~\bibnamefont{Lee}},\ and\ \bibinfo {author}
  {\bibfnamefont{G.}~\bibnamefont{Chong}},\ }%
  \bibfield{journal}{%
  \Doi{10.1103/PhysRevA.70.053621}{\bibinfo {journal} {Phys. Rev. A}}\ }%
  \textbf{\bibinfo {volume} {70}},\ \bibinfo {pages} {053621} (\bibinfo {year}
  {2004})%
  \bibAnnoteFile{NoStop}{HaiEtAl2004}%
\bibitem{MartinRuostekoski2012}%
  \BibitemOpen
  \bibfield{author}{%
  \bibinfo {author} {\bibfnamefont{A.~D.}\ \bibnamefont{Martin}}\ and\ \bibinfo
  {author} {\bibfnamefont{J.}~\bibnamefont{Ruostekoski}},\ }%
  \bibfield{journal}{%
  \Doi{10.1088/1367-2630/14/4/043040}{\bibinfo {journal} {New J. Phys.}}\ }%
  \textbf{\bibinfo {volume} {14}},\ \bibinfo {pages} {043040} (\bibinfo {year}
  {2012})%
  \bibAnnoteFile{NoStop}{MartinRuostekoski2012}%
\bibitem{HelmEtAl2012}%
  \BibitemOpen
  \bibfield{author}{%
  \bibinfo {author} {\bibfnamefont{J.~L.}\ \bibnamefont{Helm}}, \bibinfo
  {author} {\bibfnamefont{T.~P.}\ \bibnamefont{Billam}},\ and\ \bibinfo
  {author} {\bibfnamefont{S.~A.}\ \bibnamefont{Gardiner}},\ }%
  \bibfield{journal}{%
  \Doi{10.1103/PhysRevA.85.053621}{\bibinfo {journal} {Phys. Rev. A}}\ }%
  \textbf{\bibinfo {volume} {85}},\ \bibinfo {pages} {053621} (\bibinfo {year}
  {2012})%
  \bibAnnoteFile{NoStop}{HelmEtAl2012}%
\bibitem{CuevasEtAl2013}%
  \BibitemOpen
  \bibfield{author}{%
  \bibinfo {author} {\bibfnamefont{J.}~\bibnamefont{Cuevas}}, \bibinfo {author}
  {\bibfnamefont{P.~G.}\ \bibnamefont{Kevrekidis}}, \bibinfo {author}
  {\bibfnamefont{B.~A.}\ \bibnamefont{Malomed}}, \bibinfo {author}
  {\bibfnamefont{P.}~\bibnamefont{Dyke}},\ and\ \bibinfo {author}
  {\bibfnamefont{R.~G.}\ \bibnamefont{Hulet}},\ }%
  \bibfield{journal}{%
  \Doi{10.1088/1367-2630/15/6/063006}{\bibinfo {journal} {New J. Phys.}}\ }%
  \textbf{\bibinfo {volume} {15}},\ \bibinfo {pages} {063006} (\bibinfo {year}
  {2013})%
  \bibAnnoteFile{NoStop}{CuevasEtAl2013}%
\bibitem{PoloAhufinger2013}%
  \BibitemOpen
  \bibfield{author}{%
  \bibinfo {author} {\bibfnamefont{J.}~\bibnamefont{Polo}}\ and\ \bibinfo
  {author} {\bibfnamefont{V.}~\bibnamefont{Ahufinger}},\ }%
  \bibfield{journal}{%
  \Doi{10.1103/PhysRevA.88.053628}{\bibinfo {journal} {Phys. Rev. A}}\ }%
  \textbf{\bibinfo {volume} {88}},\ \bibinfo {pages} {053628} (\bibinfo {year}
  {2013})%
  \bibAnnoteFile{NoStop}{PoloAhufinger2013}%
\bibitem{LaiHaus1989}%
  \BibitemOpen
  \bibfield{author}{%
  \bibinfo {author} {\bibfnamefont{Y.}~\bibnamefont{Lai}}\ and\ \bibinfo
  {author} {\bibfnamefont{H.~A.}\ \bibnamefont{Haus}},\ }%
  \bibfield{journal}{%
  \Doi{10.1103/PhysRevA.40.854}{\bibinfo {journal} {Phys. Rev. A}}\ }%
  \textbf{\bibinfo {volume} {40}},\ \bibinfo {pages} {854} (\bibinfo {year}
  {1989})%
  \bibAnnoteFile{NoStop}{LaiHaus1989}%
\bibitem{DrummondEtAl1993}%
  \BibitemOpen
  \bibfield{author}{%
  \bibinfo {author} {\bibfnamefont{P.~D.}\ \bibnamefont{{Drummond}}}, \bibinfo
  {author} {\bibfnamefont{R.~M.}\ \bibnamefont{{Shelby}}}, \bibinfo {author}
  {\bibfnamefont{S.~R.}\ \bibnamefont{{Friberg}}},\ and\ \bibinfo {author}
  {\bibfnamefont{Y.}~\bibnamefont{{Yamamoto}}},\ }%
  \bibfield{journal}{%
  \Doi{10.1038/365307a0}{\bibinfo {journal} {Nature (London)}}\ }%
  \textbf{\bibinfo {volume} {365}},\ \bibinfo {pages} {307} (\bibinfo {year}
  {1993})%
  \bibAnnoteFile{NoStop}{DrummondEtAl1993}%
\bibitem{CarrBrand2004}%
  \BibitemOpen
  \bibfield{author}{%
  \bibinfo {author} {\bibfnamefont{L.~D.}\ \bibnamefont{Carr}}\ and\ \bibinfo
  {author} {\bibfnamefont{J.}~\bibnamefont{Brand}},\ }%
  \bibfield{journal}{%
  \Doi{10.1103/PhysRevLett.92.040401}{\bibinfo {journal} {Phys. Rev. Lett.}}\
  }%
  \textbf{\bibinfo {volume} {92}},\ \bibinfo {pages} {040401} (\bibinfo {year}
  {2004})%
  \bibAnnoteFile{NoStop}{CarrBrand2004}%
\bibitem{StreltsovEtAl2011}%
  \BibitemOpen
  \bibfield{author}{%
  \bibinfo {author} {\bibfnamefont{A.~I.}\ \bibnamefont{Streltsov}}, \bibinfo
  {author} {\bibfnamefont{O.~E.}\ \bibnamefont{Alon}},\ and\ \bibinfo {author}
  {\bibfnamefont{L.~S.}\ \bibnamefont{Cederbaum}},\ }%
  \bibfield{journal}{%
  \Doi{10.1103/PhysRevLett.106.240401}{\bibinfo {journal} {Phys. Rev. Lett.}}\
  }%
  \textbf{\bibinfo {volume} {106}},\ \bibinfo {pages} {240401} (\bibinfo {year}
  {2011})%
  \bibAnnoteFile{NoStop}{StreltsovEtAl2011}%
\bibitem{FogartyEtAl2013}%
  \BibitemOpen
  \bibfield{author}{%
  \bibinfo {author} {\bibfnamefont{T.}~\bibnamefont{{Fogarty}}}, \bibinfo
  {author} {\bibfnamefont{A.}~\bibnamefont{{Kiely}}}, \bibinfo {author}
  {\bibfnamefont{S.}~\bibnamefont{{Campbell}}},\ and\ \bibinfo {author}
  {\bibfnamefont{T.}~\bibnamefont{{Busch}}},\ }%
  \bibfield{journal}{%
  \Doi{10.1103/PhysRevA.87.043630}{\bibinfo {journal} {Phys.\ Rev.\ A}}\ }%
  \textbf{\bibinfo {volume} {87}},\ \bibinfo {eid} {043630} (\bibinfo {year}
  {2013})%
  \bibAnnoteFile{NoStop}{FogartyEtAl2013}%
\bibitem{DelandeEtAl2013}%
  \BibitemOpen
  \bibfield{author}{%
  \bibinfo {author} {\bibfnamefont{D.}~\bibnamefont{{Delande}}}, \bibinfo
  {author} {\bibfnamefont{K.}~\bibnamefont{{Sacha}}}, \bibinfo {author}
  {\bibfnamefont{M.}~\bibnamefont{{P{\l}odzie{\'n}}}}, \bibinfo {author}
  {\bibfnamefont{S.~K.}\ \bibnamefont{{Avazbaev}}},\ and\ \bibinfo {author}
  {\bibfnamefont{J.}~\bibnamefont{{Zakrzewski}}},\ }%
  \bibfield{journal}{%
  \Doi{10.1088/1367-2630/15/4/045021}{\bibinfo {journal} {New J. Phys.}}\ }%
  \textbf{\bibinfo {volume} {15}},\ \bibinfo {eid} {045021} (\bibinfo {year}
  {2013})%
  \bibAnnoteFile{NoStop}{DelandeEtAl2013}%
\bibitem{GertjerenkenEtAl2013}%
  \BibitemOpen
  \bibfield{author}{%
  \bibinfo {author} {\bibfnamefont{B.}~\bibnamefont{Gertjerenken}}, \bibinfo
  {author} {\bibfnamefont{T.~P.}\ \bibnamefont{Billam}}, \bibinfo {author}
  {\bibfnamefont{C.~L.}\ \bibnamefont{Blackley}}, \bibinfo {author}
  {\bibfnamefont{C.~R.}\ \bibnamefont{Le~Sueur}}, \bibinfo {author}
  {\bibfnamefont{L.}~\bibnamefont{Khaykovich}}, \bibinfo {author}
  {\bibfnamefont{S.~L.}\ \bibnamefont{Cornish}},\ and\ \bibinfo {author}
  {\bibfnamefont{C.}~\bibnamefont{Weiss}},\ }%
  \bibfield{journal}{%
  \Doi{10.1103/PhysRevLett.111.100406}{\bibinfo {journal} {Phys. Rev. Lett.}}\
  }%
  \textbf{\bibinfo {volume} {111}},\ \bibinfo {pages} {100406} (\bibinfo {year}
  {2013})%
  \bibAnnoteFile{NoStop}{GertjerenkenEtAl2013}%
\bibitem{BarbieroSalasnich2014}%
  \BibitemOpen
  \bibfield{author}{%
  \bibinfo {author} {\bibfnamefont{L.}~\bibnamefont{{Barbiero}}}\ and\ \bibinfo
  {author} {\bibfnamefont{L.}~\bibnamefont{{Salasnich}}},\ }%
  \bibfield{journal}{%
  \Doi{10.1103/PhysRevA.89.063605}{\bibinfo {journal} {Phys. Rev. A}}\ }%
  \textbf{\bibinfo {volume} {89}},\ \bibinfo {pages} {063605} (\bibinfo {year}
  {2014})%
  \bibAnnoteFile{NoStop}{BarbieroSalasnich2014}%
\bibitem{McGuire1964}%
  \BibitemOpen
  \bibfield{author}{%
  \bibinfo {author} {\bibfnamefont{J.~B.}\ \bibnamefont{McGuire}},\ }%
  \bibfield{journal}{%
  \Doi{10.1063/1.1704156}{\bibinfo {journal} {J. Math. Phys.}}\ }%
  \textbf{\bibinfo {volume} {5}},\ \bibinfo {pages} {622} (\bibinfo {year}
  {1964})%
  \bibAnnoteFile{NoStop}{McGuire1964}%
\bibitem{CalogeroDegasperis1975}%
  \BibitemOpen
  \bibfield{author}{%
  \bibinfo {author} {\bibfnamefont{F.}~\bibnamefont{Calogero}}\ and\ \bibinfo
  {author} {\bibfnamefont{A.}~\bibnamefont{Degasperis}},\ }%
  \bibfield{journal}{%
  \Doi{10.1103/PhysRevA.11.265}{\bibinfo {journal} {Phys. Rev. A}}\ }%
  \textbf{\bibinfo {volume} {11}},\ \bibinfo {pages} {265} (\bibinfo {year}
  {1975})%
  \bibAnnoteFile{NoStop}{CalogeroDegasperis1975}%
\bibitem{CastinHerzog2001}%
  \BibitemOpen
  \bibfield{author}{%
  \bibinfo {author} {\bibfnamefont{Y.}~\bibnamefont{Castin}}\ and\ \bibinfo
  {author} {\bibfnamefont{C.}~\bibnamefont{Herzog}},\ }%
  \bibfield{journal}{%
  \Doi{10.1016/S1296-2147(01)01183-0}{\bibinfo {journal} {C. R. Acad. Sci.
  Paris, Ser. IV}}\ }%
  \textbf{\bibinfo {volume} {2}},\ \bibinfo {pages} {419} (\bibinfo {year}
  {2001}),\
  \Eprint{http://arxiv.org/abs/arXiv:cond-mat/0012040}{arXiv:cond-mat/0012040}%
  \bibAnnoteFile{NoStop}{CastinHerzog2001}%
\bibitem{LiebLiniger1963}%
  \BibitemOpen
  \bibfield{author}{%
  \bibinfo {author} {\bibfnamefont{E.~H.}\ \bibnamefont{Lieb}}\ and\ \bibinfo
  {author} {\bibfnamefont{W.}~\bibnamefont{Liniger}},\ }%
  \bibfield{journal}{%
  \Doi{10.1103/PhysRev.130.1605}{\bibinfo {journal} {Phys. Rev.}}\ }%
  \textbf{\bibinfo {volume} {130}},\ \bibinfo {pages} {1605} (\bibinfo {year}
  {1963})%
  \bibAnnoteFile{NoStop}{LiebLiniger1963}%
\bibitem{MazetsKurizki2006}%
  \BibitemOpen
  \bibfield{author}{%
  \bibinfo {author} {\bibfnamefont{I.~E.}\ \bibnamefont{Mazets}}\ and\ \bibinfo
  {author} {\bibfnamefont{G.}~\bibnamefont{Kurizki}},\ }%
  \bibfield{journal}{%
  \Doi{10.1209/epl/i2006-10260-0}{\bibinfo {journal} {EPL (Europhys.\ Lett.)}}\
  }%
  \textbf{\bibinfo {volume} {76}},\ \bibinfo {pages} {196} (\bibinfo {year}
  {2006})%
  \bibAnnoteFile{NoStop}{MazetsKurizki2006}%
\bibitem{UhlenbeckOrnstein1930}%
  \BibitemOpen
  \bibfield{author}{%
  \bibinfo {author} {\bibfnamefont{G.~E.}\ \bibnamefont{{Uhlenbeck}}}\ and\
  \bibinfo {author} {\bibfnamefont{L.~S.}\ \bibnamefont{{Ornstein}}},\ }%
  \bibfield{journal}{%
  \Doi{10.1103/PhysRev.36.823}{\bibinfo {journal} {Phys. Rev.}}\ }%
  \textbf{\bibinfo {volume} {36}},\ \bibinfo {pages} {823} (\bibinfo {month}
  {Sep.}\ \bibinfo {year} {1930})%
  \bibAnnoteFile{NoStop}{UhlenbeckOrnstein1930}%
\bibitem{GrabertEtAl1988}%
  \BibitemOpen
  \bibfield{author}{%
  \bibinfo {author} {\bibfnamefont{H.}~\bibnamefont{{Grabert}}}, \bibinfo
  {author} {\bibfnamefont{P.}~\bibnamefont{{Schramm}}},\ and\ \bibinfo {author}
  {\bibfnamefont{G.-L.}\ \bibnamefont{{Ingold}}},\ }%
  \bibfield{journal}{%
  \Doi{10.1016/0370-1573(88)90023-3}{\bibinfo {journal} {Phys. Rep.}}\ }%
  \textbf{\bibinfo {volume} {168}},\ \bibinfo {pages} {115} (\bibinfo {month}
  {Oct.}\ \bibinfo {year} {1988})%
  \bibAnnoteFile{NoStop}{GrabertEtAl1988}%
\bibitem{JungHanggi1991}%
  \BibitemOpen
  \bibfield{author}{%
  \bibinfo {author} {\bibfnamefont{P.}~\bibnamefont{{Jung}}}\ and\ \bibinfo
  {author} {\bibfnamefont{P.}~\bibnamefont{{H{\"a}nggi}}},\ }%
  \bibfield{journal}{%
  \Doi{10.1103/PhysRevA.44.8032}{\bibinfo {journal} {\pra}}\ }%
  \textbf{\bibinfo {volume} {44}},\ \bibinfo {pages} {8032} (\bibinfo {year}
  {1991})%
  \bibAnnoteFile{NoStop}{JungHanggi1991}%
\bibitem{WangEtAl2002}%
  \BibitemOpen
  \bibfield{author}{%
  \bibinfo {author} {\bibfnamefont{G.~M.}\ \bibnamefont{{Wang}}}, \bibinfo
  {author} {\bibfnamefont{E.~M.}\ \bibnamefont{{Sevick}}}, \bibinfo {author}
  {\bibfnamefont{E.}~\bibnamefont{{Mittag}}}, \bibinfo {author}
  {\bibfnamefont{D.~J.}\ \bibnamefont{{Searles}}},\ and\ \bibinfo {author}
  {\bibfnamefont{D.~J.}\ \bibnamefont{{Evans}}},\ }%
  \bibfield{journal}{%
  \Doi{10.1103/PhysRevLett.89.050601}{\bibinfo {journal} {Phys. Rev. Lett.}}\
  }%
  \textbf{\bibinfo {volume} {89}},\ \bibinfo {eid} {050601} (\bibinfo {year}
  {2002})%
  \bibAnnoteFile{NoStop}{WangEtAl2002}%
\bibitem{LukicEtAl2005}%
  \BibitemOpen
  \bibfield{author}{%
  \bibinfo {author} {\bibfnamefont{B.}~\bibnamefont{Luki\ifmmode~\acute{c}\else
  \'{c}\fi{}}}, \bibinfo {author} {\bibfnamefont{S.}~\bibnamefont{Jeney}},
  \bibinfo {author} {\bibfnamefont{C.}~\bibnamefont{Tischer}}, \bibinfo
  {author} {\bibfnamefont{A.~J.}\ \bibnamefont{Kulik}}, \bibinfo {author}
  {\bibfnamefont{L.}~\bibnamefont{Forr\'o}},\ and\ \bibinfo {author}
  {\bibfnamefont{E.-L.}\ \bibnamefont{Florin}},\ }%
  \bibfield{journal}{%
  \Doi{10.1103/PhysRevLett.95.160601}{\bibinfo {journal} {Phys. Rev. Lett.}}\
  }%
  \textbf{\bibinfo {volume} {95}},\ \bibinfo {pages} {160601} (\bibinfo {year}
  {2005})%
  \bibAnnoteFile{NoStop}{LukicEtAl2005}%
\bibitem{KoepplEtAl2006}%
  \BibitemOpen
  \bibfield{author}{%
  \bibinfo {author} {\bibfnamefont{M.}~\bibnamefont{K\"oppl}}, \bibinfo
  {author} {\bibfnamefont{P.}~\bibnamefont{Henseler}}, \bibinfo {author}
  {\bibfnamefont{A.}~\bibnamefont{Erbe}}, \bibinfo {author}
  {\bibfnamefont{P.}~\bibnamefont{Nielaba}},\ and\ \bibinfo {author}
  {\bibfnamefont{P.}~\bibnamefont{Leiderer}},\ }%
  \bibfield{journal}{%
  \Doi{10.1103/PhysRevLett.97.208302}{\bibinfo {journal} {Phys. Rev. Lett.}}\
  }%
  \textbf{\bibinfo {volume} {97}},\ \bibinfo {pages} {208302} (\bibinfo {year}
  {2006})%
  \bibAnnoteFile{NoStop}{KoepplEtAl2006}%
\bibitem{AndoSkolnick2010}%
  \BibitemOpen
  \bibfield{author}{%
  \bibinfo {author} {\bibfnamefont{T.}~\bibnamefont{Ando}}\ and\ \bibinfo
  {author} {\bibfnamefont{J.}~\bibnamefont{Skolnick}},\ }%
  \bibfield{journal}{%
  \Doi{10.1073/pnas.1011354107}{\bibinfo {journal} {Proc. Natl. Acad. Sci.}}\
  }%
  \textbf{\bibinfo {volume} {107}},\ \bibinfo {pages} {18457} (\bibinfo {year}
  {2010})%
  \bibAnnoteFile{NoStop}{AndoSkolnick2010}%
\bibitem{SteinigewegEtAl2007}%
  \BibitemOpen
  \bibfield{author}{%
  \bibinfo {author} {\bibfnamefont{R.}~\bibnamefont{Steinigeweg}}, \bibinfo
  {author} {\bibfnamefont{H.-P.}\ \bibnamefont{Breuer}},\ and\ \bibinfo
  {author} {\bibfnamefont{J.}~\bibnamefont{Gemmer}},\ }%
  \bibfield{journal}{%
  \Doi{10.1103/PhysRevLett.99.150601}{\bibinfo {journal} {Phys.\ Rev.\ Lett.}}\
  }%
  \textbf{\bibinfo {volume} {99}},\ \bibinfo {pages} {150601} (\bibinfo {year}
  {2007})%
  \bibAnnoteFile{NoStop}{SteinigewegEtAl2007}%
\bibitem{Metzler2000}%
  \BibitemOpen
  \bibfield{author}{%
  \bibinfo {author} {\bibfnamefont{R.}~\bibnamefont{Metzler}}\ and\ \bibinfo
  {author} {\bibfnamefont{J.}~\bibnamefont{Klafter}},\ }%
  \bibfield{journal}{%
  \Doi{http://dx.doi.org/10.1016/S0370-1573(00)00070-3}{\bibinfo {journal}
  {Phys. Rep.}}\ }%
  \textbf{\bibinfo {volume} {339}},\ \bibinfo {pages} {1} (\bibinfo {year}
  {2000})%
  \bibAnnoteFile{NoStop}{Metzler2000}%
\bibitem{SiemsEtAl2012}%
  \BibitemOpen
  \bibfield{author}{%
  \bibinfo {author} {\bibfnamefont{U.}~\bibnamefont{{Siems}}}, \bibinfo
  {author} {\bibfnamefont{C.}~\bibnamefont{{Kreuter}}}, \bibinfo {author}
  {\bibfnamefont{A.}~\bibnamefont{{Erbe}}}, \bibinfo {author}
  {\bibfnamefont{N.}~\bibnamefont{{Schwierz}}}, \bibinfo {author}
  {\bibfnamefont{S.}~\bibnamefont{{Sengupta}}}, \bibinfo {author}
  {\bibfnamefont{P.}~\bibnamefont{{Leiderer}}},\ and\ \bibinfo {author}
  {\bibfnamefont{P.}~\bibnamefont{{Nielaba}}},\ }%
  \bibfield{journal}{%
  \Doi{10.1038/srep01015}{\bibinfo {journal} {Sci. Rep.}}\ }%
  \textbf{\bibinfo {volume} {2}},\ \bibinfo {pages} {1015} (\bibinfo {year}
  {2012})%
  \bibAnnoteFile{NoStop}{SiemsEtAl2012}%
\bibitem{TurivEtAl2013}%
  \BibitemOpen
  \bibfield{author}{%
  \bibinfo {author} {\bibfnamefont{T.}~\bibnamefont{{Turiv}}}, \bibinfo
  {author} {\bibfnamefont{I.}~\bibnamefont{{Lazo}}}, \bibinfo {author}
  {\bibfnamefont{A.}~\bibnamefont{{Brodin}}}, \bibinfo {author}
  {\bibfnamefont{B.~I.}\ \bibnamefont{{Lev}}}, \bibinfo {author}
  {\bibfnamefont{V.}~\bibnamefont{{Reiffenrath}}}, \bibinfo {author}
  {\bibfnamefont{V.~G.}\ \bibnamefont{{Nazarenko}}},\ and\ \bibinfo {author}
  {\bibfnamefont{O.~D.}\ \bibnamefont{{Lavrentovich}}},\ }%
  \bibfield{journal}{%
  \Doi{10.1126/science.1240591}{\bibinfo {journal} {Science}}\ }%
  \textbf{\bibinfo {volume} {342}},\ \bibinfo {pages} {1351} (\bibinfo {year}
  {2013})%
  \bibAnnoteFile{NoStop}{TurivEtAl2013}%
\bibitem{MetzlerKlafter2004}%
  \BibitemOpen
  \bibfield{author}{%
  \bibinfo {author} {\bibfnamefont{R.}~\bibnamefont{Metzler}}\ and\ \bibinfo
  {author} {\bibfnamefont{J.}~\bibnamefont{Klafter}},\ }%
  \bibfield{journal}{%
  \Doi{doi:10.1088/0305-4470/37/31/R01}{\bibinfo {journal} {J. Phys. A: Math.
  Gen.}}\ }%
  \textbf{\bibinfo {volume} {37}},\ \bibinfo {pages} {R161} (\bibinfo {year}
  {2004})%
  \bibAnnoteFile{NoStop}{MetzlerKlafter2004}%
\bibitem{DriesEtAl2010}%
  \BibitemOpen
  \bibfield{author}{%
  \bibinfo {author} {\bibfnamefont{D.}~\bibnamefont{{Dries}}}, \bibinfo
  {author} {\bibfnamefont{S.~E.}\ \bibnamefont{{Pollack}}}, \bibinfo {author}
  {\bibfnamefont{J.~M.}\ \bibnamefont{{Hitchcock}}},\ and\ \bibinfo {author}
  {\bibfnamefont{R.~G.}\ \bibnamefont{{Hulet}}},\ }%
  \bibfield{journal}{%
  \Doi{10.1103/PhysRevA.82.033603}{\bibinfo {journal} {\pra}}\ }%
  \textbf{\bibinfo {volume} {82}},\ \bibinfo {eid} {033603} (\bibinfo {year}
  {2010})%
  \bibAnnoteFile{NoStop}{DriesEtAl2010}%
\bibitem{SinhaEtAl2006}%
  \BibitemOpen
  \bibfield{author}{%
  \bibinfo {author} {\bibfnamefont{S.}~\bibnamefont{Sinha}}, \bibinfo {author}
  {\bibfnamefont{A.~Y.}\ \bibnamefont{Cherny}}, \bibinfo {author}
  {\bibfnamefont{D.}~\bibnamefont{Kovrizhin}},\ and\ \bibinfo {author}
  {\bibfnamefont{J.}~\bibnamefont{Brand}},\ }%
  \bibfield{journal}{%
  \Doi{10.1103/PhysRevLett.96.030406}{\bibinfo {journal} {Phys. Rev. Lett.}}\
  }%
  \textbf{\bibinfo {volume} {96}},\ \bibinfo {pages} {030406} (\bibinfo {year}
  {2006})%
  \bibAnnoteFile{NoStop}{SinhaEtAl2006}%
\bibitem{WeissCastin2009}%
  \BibitemOpen
  \bibfield{author}{%
  \bibinfo {author} {\bibfnamefont{C.}~\bibnamefont{Weiss}}\ and\ \bibinfo
  {author} {\bibfnamefont{Y.}~\bibnamefont{Castin}},\ }%
  \bibfield{journal}{%
  \Doi{10.1103/PhysRevLett.102.010403}{\bibinfo {journal} {Phys.\ Rev.\
  Lett.}}\ }%
  \textbf{\bibinfo {volume} {102}},\ \bibinfo {pages} {010403} (\bibinfo {year}
  {2009})%
  \bibAnnoteFile{NoStop}{WeissCastin2009}%
\bibitem{GrimmEtAl2000}%
  \BibitemOpen
  \bibfield{author}{%
  \bibinfo {author} {\bibfnamefont{R.}~\bibnamefont{{Grimm}}}, \bibinfo
  {author} {\bibfnamefont{M.}~\bibnamefont{{Weidem{\"u}ller}}},\ and\ \bibinfo
  {author} {\bibfnamefont{Y.~B.}\ \bibnamefont{{Ovchinnikov}}},\ }%
  \bibfield{journal}{%
  \Doi{10.1016/S1049-250X(08)60186-X}{\bibinfo {journal} {Adv. At. Mol. Opt.
  Phys.}}\ }%
  \textbf{\bibinfo {volume} {42}},\ \bibinfo {pages} {95} (\bibinfo {year}
  {2000}),\ \Eprint{http://arxiv.org/abs/physics/9902072}{physics/9902072}%
  \bibAnnoteFile{NoStop}{GrimmEtAl2000}%
\bibitem{AndersonEtAl1995}%
  \BibitemOpen
  \bibfield{author}{%
  \bibinfo {author} {\bibfnamefont{M.~H.}\ \bibnamefont{Anderson}}, \bibinfo
  {author} {\bibfnamefont{J.~R.}\ \bibnamefont{Ensher}}, \bibinfo {author}
  {\bibfnamefont{M.~R.}\ \bibnamefont{Matthews}}, \bibinfo {author}
  {\bibfnamefont{C.~E.}\ \bibnamefont{Wieman}},\ and\ \bibinfo {author}
  {\bibfnamefont{E.~A.}\ \bibnamefont{Cornell}},\ }%
  \bibfield{journal}{%
  \Doi{10.1126/science.269.5221.198}{\bibinfo {journal} {Science}}\ }%
  \textbf{\bibinfo {volume} {269}},\ \bibinfo {pages} {198} (\bibinfo {year}
  {1995})%
  \bibAnnoteFile{NoStop}{AndersonEtAl1995}%
\bibitem{Davis1993}%
  \BibitemOpen
  \bibfield{author}{%
  \bibinfo {author} {\bibfnamefont{M.~H.~A.}\ \bibnamefont{Davis}},\ }%
  \emph{\bibinfo {title} {Markov models and optimization}}\ (\bibinfo
  {publisher} {Chapman {\&} Hall},\ \bibinfo {address} {London},\ \bibinfo
  {year} {1993})%
  \bibAnnoteFile{NoStop}{Davis1993}%
\bibitem{Olshanii1998}%
  \BibitemOpen
  \bibfield{author}{%
  \bibinfo {author} {\bibfnamefont{M.}~\bibnamefont{Olshanii}},\ }%
  \bibfield{journal}{%
  \Doi{10.1103/PhysRevLett.81.938}{\bibinfo {journal} {Phys. Rev. Lett.}}\ }%
  \textbf{\bibinfo {volume} {81}},\ \bibinfo {pages} {938} (\bibinfo {year}
  {1998})%
  \bibAnnoteFile{NoStop}{Olshanii1998}%
\bibitem{Castin2009}%
  \BibitemOpen
  \bibfield{author}{%
  \bibinfo {author} {\bibfnamefont{Y.}~\bibnamefont{Castin}},\ }%
  \bibfield{journal}{%
  \Doi{10.1140/epjb/e2008-00407-3}{\bibinfo {journal} {Eur. Phys. J. B}}\ }%
  \textbf{\bibinfo {volume} {68}},\ \bibinfo {pages} {317} (\bibinfo {year}
  {2009})%
  \bibAnnoteFile{NoStop}{Castin2009}%
\bibitem{BonitzEtAl2007}%
  \BibitemOpen
  \bibfield{author}{%
  \bibinfo {author} {\bibfnamefont{M.}~\bibnamefont{Bonitz}}, \bibinfo {author}
  {\bibfnamefont{K.}~\bibnamefont{Balzer}},\ and\ \bibinfo {author}
  {\bibfnamefont{R.}~\bibnamefont{van Leeuwen}},\ }%
  \bibfield{journal}{%
  \Doi{10.1103/PhysRevB.76.045341}{\bibinfo {journal} {Phys. Rev. B}}\ }%
  \textbf{\bibinfo {volume} {76}},\ \bibinfo {pages} {045341} (\bibinfo {year}
  {2007})%
  \bibAnnoteFile{NoStop}{BonitzEtAl2007}%
\bibitem{Gertjerenken2013}%
  \BibitemOpen
  \bibfield{author}{%
  \bibinfo {author} {\bibfnamefont{B.}~\bibnamefont{Gertjerenken}},\ }%
  \bibfield{journal}{%
  \Doi{10.1103/PhysRevA.88.053623}{\bibinfo {journal} {Phys. Rev. A}}\ }%
  \textbf{\bibinfo {volume} {88}},\ \bibinfo {pages} {053623} (\bibinfo {year}
  {2013})%
  \bibAnnoteFile{NoStop}{Gertjerenken2013}%
\bibitem{Fluegge1990}%
  \BibitemOpen
  \bibfield{author}{%
  \bibinfo {author} {\bibfnamefont{S.}~\bibnamefont{{Fl\"ugge}}},\ }%
  \emph{\bibinfo {title} {Rechenmethoden der Quantentheorie}}\ (\bibinfo
  {publisher} {Springer},\ \bibinfo {address} {Berlin},\ \bibinfo {year}
  {1990})%
  \bibAnnoteFile{NoStop}{Fluegge1990}%
\bibitem{Lieb2002}%
  \BibitemOpen
  \bibfield{author}{%
  \bibinfo {author} {\bibfnamefont{E.~H.}\ \bibnamefont{Lieb}}, \bibinfo
  {author} {\bibfnamefont{R.}~\bibnamefont{Seiringer}}, \bibinfo {author}
  {\bibfnamefont{J.~P.}\ \bibnamefont{Solovej}},\ and\ \bibinfo {author}
  {\bibfnamefont{J.}~\bibnamefont{Yngvason}},\ }%
  \bibfield{journal}{%
  \bibinfo {journal} {Current Developments in Mathematics, 2001},\ \bibinfo
  {pages} {131}}%
   (\bibinfo {year} {2002})%
  \bibAnnoteFile{NoStop}{Lieb2002}%
\bibitem{GertjerenkenWeiss2013}%
  \BibitemOpen
  \bibfield{author}{%
  \bibinfo {author} {\bibfnamefont{B.}~\bibnamefont{{Gertjerenken}}}\ and\
  \bibinfo {author} {\bibfnamefont{C.}~\bibnamefont{{Weiss}}},\ }%
  \bibfield{journal}{%
  \Doi{10.1103/PhysRevA.88.033608}{\bibinfo {journal} {\pra}}\ }%
  \textbf{\bibinfo {volume} {88}},\ \bibinfo {eid} {033608} (\bibinfo {year}
  {2013})%
  \bibAnnoteFile{NoStop}{GertjerenkenWeiss2013}%
\bibitem{GertjerenkenWeiss2012}%
  \BibitemOpen
  \bibfield{author}{%
  \bibinfo {author} {\bibfnamefont{B.}~\bibnamefont{Gertjerenken}}\ and\
  \bibinfo {author} {\bibfnamefont{C.}~\bibnamefont{Weiss}},\ }%
  \bibfield{journal}{%
  \Doi{10.1088/0953-4075/45/16/165301}{\bibinfo {journal} {J. Phys. B}}\ }%
  \textbf{\bibinfo {volume} {45}},\ \bibinfo {pages} {165301} (\bibinfo {year}
  {2012})%
  \bibAnnoteFile{NoStop}{GertjerenkenWeiss2012}%
\bibitem{GauntEtAl2013}%
  \BibitemOpen
  \bibfield{author}{%
  \bibinfo {author} {\bibfnamefont{A.~L.}\ \bibnamefont{Gaunt}}, \bibinfo
  {author} {\bibfnamefont{T.~F.}\ \bibnamefont{Schmidutz}}, \bibinfo {author}
  {\bibfnamefont{I.}~\bibnamefont{Gotlibovych}}, \bibinfo {author}
  {\bibfnamefont{R.~P.}\ \bibnamefont{Smith}},\ and\ \bibinfo {author}
  {\bibfnamefont{Z.}~\bibnamefont{Hadzibabic}},\ }%
  \bibfield{journal}{%
  \Doi{10.1103/PhysRevLett.110.200406}{\bibinfo {journal} {Phys. Rev. Lett.}}\
  }%
  \textbf{\bibinfo {volume} {110}},\ \bibinfo {pages} {200406} (\bibinfo {year}
  {2013})%
  \bibAnnoteFile{NoStop}{GauntEtAl2013}%
\bibitem{SchmidutzEtAl2014}%
  \BibitemOpen
  \bibfield{author}{%
  \bibinfo {author} {\bibfnamefont{T.~F.}\ \bibnamefont{Schmidutz}}, \bibinfo
  {author} {\bibfnamefont{I.}~\bibnamefont{Gotlibovych}}, \bibinfo {author}
  {\bibfnamefont{A.~L.}\ \bibnamefont{Gaunt}}, \bibinfo {author}
  {\bibfnamefont{R.~P.}\ \bibnamefont{Smith}}, \bibinfo {author}
  {\bibfnamefont{N.}~\bibnamefont{Navon}},\ and\ \bibinfo {author}
  {\bibfnamefont{Z.}~\bibnamefont{Hadzibabic}},\ }%
  \bibfield{journal}{%
  \Doi{10.1103/PhysRevLett.112.040403}{\bibinfo {journal} {Phys. Rev. Lett.}}\
  }%
  \textbf{\bibinfo {volume} {112}},\ \bibinfo {pages} {040403} (\bibinfo {year}
  {2014})%
  \bibAnnoteFile{NoStop}{SchmidutzEtAl2014}%
\bibitem{maple}%
  \BibitemOpen
  \bibinfo {note} {Computer algebra programme Maple}%
  \bibAnnoteFile{NoStop}{maple}%
\bibitem{HelmEtAl2014}%
  \BibitemOpen
  \bibfield{author}{%
  \bibinfo {author} {\bibfnamefont{J.~L.}\ \bibnamefont{Helm}}, \bibinfo
  {author} {\bibfnamefont{S.~J.}\ \bibnamefont{Rooney}}, \bibinfo {author}
  {\bibfnamefont{C.}~\bibnamefont{Weiss}},\ and\ \bibinfo {author}
  {\bibfnamefont{S.~A.}\ \bibnamefont{Gardiner}},\ }%
  \bibfield{journal}{%
  \Doi{10.1103/PhysRevA.89.033610}{\bibinfo {journal} {Phys. Rev. A}}\ }%
  \textbf{\bibinfo {volume} {89}},\ \bibinfo {pages} {033610} (\bibinfo {year}
  {2014})%
  \bibAnnoteFile{NoStop}{HelmEtAl2014}%
\bibitem{HoldawayEtAl2012}%
  \BibitemOpen
  \bibfield{author}{%
  \bibinfo {author} {\bibfnamefont{D.~I.~H.}\ \bibnamefont{Holdaway}}, \bibinfo
  {author} {\bibfnamefont{C.}~\bibnamefont{Weiss}},\ and\ \bibinfo {author}
  {\bibfnamefont{S.~A.}\ \bibnamefont{Gardiner}},\ }%
  \bibfield{journal}{%
  \Doi{10.1103/PhysRevA.85.053618}{\bibinfo {journal} {Phys. Rev. A}}\ }%
  \textbf{\bibinfo {volume} {85}},\ \bibinfo {pages} {053618} (\bibinfo {year}
  {2012})%
  \bibAnnoteFile{NoStop}{HoldawayEtAl2012}%
\bibitem{ShotanEtAl2014}%
  \BibitemOpen
  \bibfield{author}{%
  \bibinfo {author} {\bibfnamefont{Z.}~\bibnamefont{Shotan}}, \bibinfo {author}
  {\bibfnamefont{O.}~\bibnamefont{Machtey}}, \bibinfo {author}
  {\bibfnamefont{S.}~\bibnamefont{Kokkelmans}},\ and\ \bibinfo {author}
  {\bibfnamefont{L.}~\bibnamefont{Khaykovich}},\ }%
  \bibfield{journal}{%
  \Doi{10.1103/PhysRevLett.113.053202}{\bibinfo {journal} {Phys. Rev. Lett.}}\
  }%
  \textbf{\bibinfo {volume} {113}},\ \bibinfo {pages} {053202} (\bibinfo {year}
  {2014})%
  \bibAnnoteFile{NoStop}{ShotanEtAl2014}%
\bibitem{DalibardEtAl1992}%
  \BibitemOpen
  \bibfield{author}{%
  \bibinfo {author} {\bibfnamefont{J.}~\bibnamefont{Dalibard}}, \bibinfo
  {author} {\bibfnamefont{Y.}~\bibnamefont{Castin}},\ and\ \bibinfo {author}
  {\bibfnamefont{K.}~\bibnamefont{M{\o}lmer}},\ }%
  \bibfield{journal}{%
  \Doi{10.1103/PhysRevLett.68.580}{\bibinfo {journal} {Phys.\ Rev.\ Lett.}}\ }%
  \textbf{\bibinfo {volume} {68}},\ \bibinfo {pages} {580} (\bibinfo {year}
  {1992})%
  \bibAnnoteFile{NoStop}{DalibardEtAl1992}%
\bibitem{DumEtAl1992}%
  \BibitemOpen
  \bibfield{author}{%
  \bibinfo {author} {\bibfnamefont{R.}~\bibnamefont{Dum}}, \bibinfo {author}
  {\bibfnamefont{P.}~\bibnamefont{Zoller}},\ and\ \bibinfo {author}
  {\bibfnamefont{H.}~\bibnamefont{Ritsch}},\ }%
  \bibfield{journal}{%
  \Doi{10.1103/PhysRevA.45.4879}{\bibinfo {journal} {Phys. Rev. A}}\ }%
  \textbf{\bibinfo {volume} {45}},\ \bibinfo {pages} {4879} (\bibinfo {year}
  {1992})%
  \bibAnnoteFile{NoStop}{DumEtAl1992}%
\bibitem{Breuer2006}%
  \BibitemOpen
  \bibfield{author}{%
  \bibinfo {author} {\bibfnamefont{H.-P.}\ \bibnamefont{Breuer}}\ and\ \bibinfo
  {author} {\bibfnamefont{F.}~\bibnamefont{Petruccione}},\ }%
  \emph{\bibinfo {title} {The Theory of Open Quantum Systems}}\ (\bibinfo
  {publisher} {Clarendon Press},\ \bibinfo {address} {Oxford},\ \bibinfo {year}
  {2006})%
  \bibAnnoteFile{NoStop}{Breuer2006}%
\bibitem{HarocheRaimond2006}%
  \BibitemOpen
  \bibfield{author}{%
  \bibinfo {author} {\bibfnamefont{S.}~\bibnamefont{Haroche}}\ and\ \bibinfo
  {author} {\bibfnamefont{J.-M.}\ \bibnamefont{Raimond}},\ }%
  \emph{\bibinfo {title} {Exploring the Quantum -- Atoms, Cavities and
  Photons}}\ (\bibinfo {publisher} {Oxford University Press},\ \bibinfo
  {address} {Oxford},\ \bibinfo {year} {2006})%
  \bibAnnoteFile{NoStop}{HarocheRaimond2006}%
\bibitem{SinatraEtAl2002}%
  \BibitemOpen
  \bibfield{author}{%
  \bibinfo {author} {\bibfnamefont{A.}~\bibnamefont{Sinatra}}, \bibinfo
  {author} {\bibfnamefont{C.}~\bibnamefont{Lobo}},\ and\ \bibinfo {author}
  {\bibfnamefont{Y.}~\bibnamefont{Castin}},\ }%
  \bibfield{journal}{%
  \Doi{10.1088/0953-4075/35/17/301}{\bibinfo {journal} {J. Phys. B}}\ }%
  \textbf{\bibinfo {volume} {35}},\ \bibinfo {pages} {3599} (\bibinfo {year}
  {2002})%
  \bibAnnoteFile{NoStop}{SinatraEtAl2002}%
\bibitem{BieniasEtAl2011}%
  \BibitemOpen
  \bibfield{author}{%
  \bibinfo {author} {\bibfnamefont{P.}~\bibnamefont{Bienias}}, \bibinfo
  {author} {\bibfnamefont{K.}~\bibnamefont{Pawlowski}}, \bibinfo {author}
  {\bibfnamefont{M.}~\bibnamefont{Gajda}},\ and\ \bibinfo {author}
  {\bibfnamefont{K.}~\bibnamefont{Rzazewski}},\ }%
  \bibfield{journal}{%
  \Doi{10.1209/0295-5075/96/10011}{\bibinfo {journal} {EPL (Europhys. Lett.)}}\
  }%
  \textbf{\bibinfo {volume} {96}},\ \bibinfo {pages} {10011} (\bibinfo {year}
  {2011})%
  \bibAnnoteFile{NoStop}{BieniasEtAl2011}%
\bibitem{LiebEtAl2000}%
  \BibitemOpen
  \bibfield{author}{%
  \bibinfo {author} {\bibfnamefont{E.~H.}\ \bibnamefont{Lieb}}, \bibinfo
  {author} {\bibfnamefont{R.}~\bibnamefont{Seiringer}},\ and\ \bibinfo {author}
  {\bibfnamefont{J.}~\bibnamefont{Yngvason}},\ }%
  \bibfield{journal}{%
  \Doi{10.1103/PhysRevA.61.043602}{\bibinfo {journal} {Phys. Rev. A}}\ }%
  \textbf{\bibinfo {volume} {61}},\ \bibinfo {pages} {043602} (\bibinfo {year}
  {2000})%
  \bibAnnoteFile{NoStop}{LiebEtAl2000}%
\bibitem{GhirardiEtAl1986}%
  \BibitemOpen
  \bibfield{author}{%
  \bibinfo {author} {\bibfnamefont{G.~C.}\ \bibnamefont{{Ghirardi}}}, \bibinfo
  {author} {\bibfnamefont{A.}~\bibnamefont{{Rimini}}},\ and\ \bibinfo {author}
  {\bibfnamefont{T.}~\bibnamefont{{Weber}}},\ }%
  \bibfield{journal}{%
  \Doi{10.1103/PhysRevD.34.470}{\bibinfo {journal} {Phys. Rev. D}}\ }%
  \textbf{\bibinfo {volume} {34}},\ \bibinfo {pages} {470} (\bibinfo {year}
  {1986})%
  \bibAnnoteFile{NoStop}{GhirardiEtAl1986}%
\bibitem{DurEtAl2002}%
  \BibitemOpen
  \bibfield{author}{%
  \bibinfo {author} {\bibfnamefont{W.}~\bibnamefont{D\"ur}}, \bibinfo {author}
  {\bibfnamefont{R.}~\bibnamefont{Raussendorf}}, \bibinfo {author}
  {\bibfnamefont{V.~M.}\ \bibnamefont{Kendon}},\ and\ \bibinfo {author}
  {\bibfnamefont{H.-J.}\ \bibnamefont{Briegel}},\ }%
  \bibfield{journal}{%
  \Doi{10.1103/PhysRevA.66.052319}{\bibinfo {journal} {Phys. Rev. A}}\ }%
  \textbf{\bibinfo {volume} {66}},\ \bibinfo {pages} {052319} (\bibinfo {year}
  {2002})%
  \bibAnnoteFile{NoStop}{DurEtAl2002}%
\bibitem{KarskiEtAl2009}%
  \BibitemOpen
  \bibfield{author}{%
  \bibinfo {author} {\bibfnamefont{M.}~\bibnamefont{Karski}}, \bibinfo {author}
  {\bibfnamefont{L.}~\bibnamefont{F{\"o}rster}}, \bibinfo {author}
  {\bibfnamefont{J.-M.}\ \bibnamefont{Choi}}, \bibinfo {author}
  {\bibfnamefont{A.}~\bibnamefont{Steffen}}, \bibinfo {author}
  {\bibfnamefont{W.}~\bibnamefont{Alt}}, \bibinfo {author}
  {\bibfnamefont{D.}~\bibnamefont{Meschede}},\ and\ \bibinfo {author}
  {\bibfnamefont{A.}~\bibnamefont{Widera}},\ }%
  \bibfield{journal}{%
  \Doi{10.1126/science.1174436}{\bibinfo {journal} {Science}}\ }%
  \textbf{\bibinfo {volume} {325}},\ \bibinfo {pages} {174} (\bibinfo {year}
  {2009})%
  \bibAnnoteFile{NoStop}{KarskiEtAl2009}%
\end{thebibliography}

\end{document}